\documentclass[12pt]{article} 
\usepackage{lmodern,eurosym,ae,tabularx,amssymb,url,epsf,epsfig,amsmath,amsfonts,graphicx,dsfont}
\font\dsrom=dsrom10 scaled 1200           
\usepackage[latin1]{inputenc}
\usepackage[english]{babel}
\usepackage{amsmath, amsfonts,amssymb}
\usepackage{amsthm}
\usepackage[T1]{fontenc}
\usepackage{graphicx}
\usepackage{geometry}
\newenvironment{pf}{\ \\ {\bf Proof: }}{\hfill\mbox{$\diamond$}\medskip}

\newtheorem{thm}{Theorem}[section]

\newtheorem{lemma}{Lemma}[section]

\newtheorem{prop}{Proposition}[section]
\newtheorem{remark}{Remark}[section]

\renewcommand{\epsilon}{\varepsilon}
\newcommand{\be}{\begin{equation*}}
\newcommand{\ben}{\begin{equation}}
\newcommand{\ee}{\end{equation*}}
\newcommand{\een}{\end{equation}}
\renewcommand{\qed}{\hfill\mbox{$\diamond$}\medskip}

\newcommand{\R}{\mathbb{R}}

\newcommand{\E}{\mathbb{E}}

\newcommand{\N}{\mathbb{N}}

\newcommand{\T}{\mathcal{T}}
\renewcommand{\phi}{\varphi}

\newcommand\ind[1]{\textrm{\dsrom{1}}_{#1}}

\begin{document}
\title{On the binomial approximation of the American put
    }
    \author{{ Damien Lamberton}\thanks{%
 Universit\'e Paris-Est, Laboratoire d'Analyse et de Math\'ematiques Appliqu\'ees (UMR 8050), UPEM, UPEC, CNRS, 
 Projet Mathrisk INRIA, F-77454, Marne-la-Vall\'ee, France - {\tt damien.lamberton@u-pem.fr}}}
    \date{This version: November, 2018}
    \maketitle
       \begin{abstract}
    We consider the binomial approximation of the American put price in the Black-Scholes model
    (with continuous dividend yield). 
    Our main result is that the error of approximation is $O((\ln n )^\alpha/n)$, where $n$ is the number of time periods and the exponent
    $\alpha$ is  a positive number, the value of which may differ according to the respective 
    levels of the interest rate and  the dividend yield.
        \end{abstract}
\section{The binomial approximation}
Consider the Black-Scholes model, in which the stock price at time $t$ is given by
\[
S_t=S_0e^{(r-d-\frac{\sigma^2}{2})t+\sigma B_t},
\]
where, under the risk-neutral probability measure, $(B_t)_{t\geq 0}$ is a standard Brownian motion.
Here, $r$ is the instantaneous interest rate, and $d$ is the dividend rate (or the foreign interest rate in the case of forex options). 
We assume $r>0$ and $d\geq 0$.

Denote by $P$ the price function of the  American put with maturity $T$ and
strike price $K$, so that
\[
P(t,x)=\sup_{\tau\in\T_{0,T-t}}\E_x\left(e^{-r\tau}f(S_\tau)\right), \quad 0\leq t\leq T, \quad x\in[0,+\infty),
\]
with $f(x)=(K-x)^+$, and $\E_x=\E\left(\cdot\;|\; S_0=x\right)$. Here, $\T_{0,t}$ denotes the set of all stopping times
with respect to the Brownian filtration, with values in the interval $[0,t]$.

For technical reasons (especially for the derivation of regularity  estimates for the second time derivative of the price function),
 it is more convenient to use the log-stock price.
So, we introduce
\[
X^x_t=x+\mu t +\sigma B_t, \quad \mbox{with } \mu=r-d-\frac{\sigma^2}{2},
\]
and
\[
U(T,x)= \sup_{\tau\in\T_{0,T}}\E\left(e^{-r\tau}\varphi(X^x_{\tau})\right),
\]
with $\phi(x)=\left(K-e^x\right)^+$. We then have
\[
P(t,x)=U(T-t,\ln(x)), \quad t>0, x>0.
\]
Note that $U(t,x)$ satisfies the following parabolic variational inequality
\[
\max\left[ -\frac{\partial U}{\partial t}+(A-r)U,\varphi-U\right]=0,
\]
with the initial condition $U(0,.)=\varphi$.

Here, $A$ is the infinitesimal generator of $X$, namely
\[
A=\frac{\sigma^2}{2}\frac{\partial ^2 }{\partial x^2}
   +\mu
               \frac{\partial }{\partial x}.
\]
Recall that, for each $T>0$, there is a real number $\tilde b(T)\leq \ln(K)$ such that
\[
U(T,x)>\varphi(x) \Leftrightarrow x>\tilde b(T).
\]
In fact, if $(b(t),0\leq t\leq T)$ is the exercise boundary of the American put with maturity $T$,
we have $\tilde b(t)=\ln(b(T-t))$.
We will also need the European value function, defined by 
\[
\bar{U}(T,x)=\E\left(e^{-r T}\varphi(X^x_{ T})\right).
\]
Note that $\bar U(0,.)=\varphi$ and 
\[
-\frac{\partial \bar U}{\partial t}+(A-r)\bar U=0.
\]
Note that, in Section~\ref{Section-estimates}, the function  $\bar U$ will be denoted by $u_\varphi$.

We now introduce the random walk approximation of Brownian motion.
To be more precise, assume $(X_n)_{n\geq 1}$ is a sequence of i.i.d.
real random variables satisfying $\E X_n^2=1$ and $\E X_n=0$, and define,
for any positive integer $n$, the process $B^{(n)}$ by
\[
 B^{(n)}_t=\displaystyle
 \sqrt{T/n}\;\displaystyle\sum_{k=1}^{[nt/T]}X_k,\quad 0\leq t\leq T,
\]
where $[nt/T]$ denotes the greatest integer in $nt/T$.

We will assume the following about the common distribution of the $X_n$'s
(cf. hypothesis (H4) of \cite{DL2002}). Note that, in the binomial case,
$X_1$ takes its values in $\{-1,+1\}$.

\begin{description}
\item[ (H4)] The random variable $X_1$ is bounded and satisfies 
$\E X_1^2=1$ and $\E X_1=\E X_1^3=0$.
\end{description}

In the following, we fix $S_0$ and set
\[
P_0=P(0,S_0)=U(T,\ln S_0).
\]
Note that, if we introduce the notation $g(x)=(K-S_0e^{\sigma x})^+$, we have
\[
P_0=\sup_{\tau\in{\cal T}_{0,T}}
      \E\left(
        e^{-r\tau}g(\mu_0\tau + B_\tau)\right),
\]
with $\mu_0=\mu/\sigma$. We now have a natural approximation of $P_0$, given by
\[
P^{(n)}_0=\sup_{\tau\in{\cal T}^{(n)}_{0,T}}
      \E\left(
        e^{-r\tau}g(\mu_0\tau + B^{(n)}_\tau)\right),
\]
where ${\cal T}^{(n)}_{0,T}$ denotes the set of all stopping times (with respect to the
natural filtration of $B^{(n)}$), with values in 
$[0,T]\cap\{0,T/n,2T/n, \ldots,(n-1)T/n,T\}$.
Our main result is the following.
\begin{thm}\label{mainTh}
 There exists a positive constant $C$ such that, for all positive integers $n$,
 \[
-C \frac{(\ln n)^{\bar\alpha}}{n} \leq P^{(n)}_0-P_0 \leq C \frac{(\ln n)^\alpha}{n},
 \]
 where $\bar\alpha=\alpha=1$ if $d>r$, and $\bar\alpha=3/2, \alpha=5/4$ if $d\leq r$.
\end{thm}
The above estimates improve our previous results (see \cite{DL2002}, Theorem 5.6) which gave an upper bound of the form 
$C \left(\frac{\sqrt{\ln n}}{n}\right)^{4/5}$. Note that, for European options, the error estimate is
$O(1/n)$ (see \cite{Diener}, \cite{Walsh}). 
We also mention the results of \cite{Liang2010} about finite difference schemes, which give the rate $O(1/\sqrt{n})$,
but their estimate is uniform over the time interval, while we concentrate on the error estimate for a fixed time.
The paper \cite{Liang2010} also has results about the approximation of the exercise boundary.
We also refer to \cite{Silvestrov2015} and its references for a review of recent results on the approximation
of American option prices.

Our approach remains the same as in \cite{DL2002}: we relate the error estimates to
the regularity of the value function. The improvement comes from a refinement of the quadratic estimates for the second order
time derivative, in the spirit of Friedman and Kinderlehrer (see \cite{Friedman1975}
and \cite{Kinderlehrer1980}). We also exploit the smoothness of the exercise boundary and its asymptotic properties close
to maturity. 

The constant $C$ in Theorem~\ref{mainTh} is related to the Berry-Esseen estimate and to the regularity of the value function.
Although it is hard to keep track of the constants in the regularity estimates, it may be worth mentioning that
they remain uniform with respect to $\mu$ and $\sigma$ as long as $(\mu,\sigma)$ remains in a compact
subset of $\R\times(0,\infty)$. A consequence of this observation is that the bounds in Theorem~\ref{mainTh} are also valid for
variants of the approximation in which the  process approximating $\ln(S_t/S_0)$, instead of being $\mu t+\sigma B^{(n)}_t$, is given by
$\mu_n t +\sigma_n  B^{(n)}_t$ at discrete times $t$, with $\mu_n=\mu+O(1/n)$ and $\sigma_n^2=\sigma^2+O(1/n)$,
as occurs in the classical risk-neutral approximation. Indeed,
standard arguments show that the value function is  locally Lipschitz-continuous  with respect to $\sigma^2$ (away from $0$) 
and $\mu$.

The paper is organized as follows. In the next Section we recall some results of \cite{DL2002}.
Section \ref{Section-estimates} is devoted to estimates for the derivatives of the value function.
The estimates are then used in Sections \ref{UB} and \ref{LB} to prove Theorem~\ref{mainTh}: in 
Section \ref{UB}, we give an upper bound for $P^{(n)}_0-P_0$ and in Section \ref{LB}, we derive the lower bound.

\bigskip

\noindent{\bf Acknowledgement: } The research on this paper has been stimulated by fruitful discussions on 
the approximation of American options with Martijn Pistorius, to whom the author is very grateful.
\section{The value function and the approximating process}

As in \cite{DL2002}, we introduce the modified value function
\[
u(t,x)=e^{-rt}U(T-t,\ln (S_0) +\mu t +\sigma x),\quad t\geq 0, \quad x\in \R.
\]
We have $P_0=u(0,0)$ and $u(T,x)=e^{-rT}U(0,\ln (S_0) +\mu T +\sigma x)=e^{-rT}(K-S_0e^{\mu T+ \sigma x})^+$ 
and, for $t\in [0,T]$, 
\begin{equation}
u(t,x)\geq e^{-rt}(K-S_0e^{\mu t+ \sigma x})^+=e^{-rt}g(\mu_0t +x).\label{*}
\end{equation}
We will need the European analogue of $u$, namely 
\[
\bar u(t,x)=e^{-rt}\bar U(T-t,\ln (S_0) +\mu t +\sigma x)=e^{-rT}\E\left(g(\mu_0T+x+B_{T-t})\right),\quad t\geq 0, x\in \R.
\]
We will also use the notation:
\[
      h=\frac{T}{n}\;.
\]
With this notation, we have
\[
B^{(n)}_t=\sqrt{h}\sum_{k=1}^{[t/h]}X_k, \quad 0\leq t\leq T.
\]
We have, for all $t\in\{0,h,2h,\ldots,(n-1)h,nh=T\}$ (cf. Proposition 3.1 of \cite{DL2002}),
   \[
     u(t, B^{(n)}_{t})=
         u(0,0)+M_{t}+\sum_{j=1}^{t/h}{\cal D}u({(j-1)h},B^{(n)}_{{(j-1)h}}),
         \]
   where $(M_t)_{0\leq t\leq T}$ is a martingale 
   (with respect to the natural filtration of $B^{(n)}$),
   such that $M_0=0$, and
\[
{\cal D}u(t,x)=\E
    \left( u\left(t+h, x+\sqrt{h}X_1\right)\right)
                      -u(t,x), \quad 0\leq t\leq T-h,\quad x\in\R.
\]
The above decomposition of $u(t, B^{(n)}_{t})$ (which is in fact Doob's decomposition)
can be viewed as a discrete version of It\^o's formula, which, for a smooth function $v: [0,T]\times \R\to \R$,
implies that $v(t,B_t)-\int_0^t\delta v(s,B_s)ds$ is a (local) martingale,
where
\[
\delta v= {\partial v\over\partial t}+{1\over 2}{\partial^2 v\over\partial x^2}.
\]
It is also easy to check that, if $v$ is smooth and 
${\cal D}v(t,x)=\E
    \left( v\left(t+h, x+\sqrt{h}X_1\right)\right)
                      -v(t,x)$, we have 
 \[
({1/ h})\times{\cal D}v(t,x)
     =\delta v(t,x)+O(h).
\]
The main technical difficulty that we have to deal with is the lack of smoothness of the modfied value function $u$.
\begin{remark}\rm\label{rem1}
  The derivatives of $u$ are related to those of $U$ by the following formulas.
  We have
  \begin{eqnarray*}
\frac{\partial u}{\partial t}(t,x)&=&e^{-rt}\left(-\frac{\partial U}{\partial t}+\mu \frac{\partial U}{\partial x} 
       -rU\right)(T-t,\ln (S_0) +\mu t +\sigma x)
    \end{eqnarray*}
       and
\begin{eqnarray*}
\frac{\partial ^2u}{\partial t^2}(t,x)&=&e^{-rt}\left(\frac{\partial ^2U}{\partial t^2}
    -2\mu \frac{\partial^2 U}{\partial t \partial x} +\mu^2\frac{\partial^2 U}{\partial x^2}\right.\\
    &&\left.
    +2r\frac{\partial U}{\partial t } -2r\mu\frac{\partial U}{\partial x }
       +r^2U\right)(T-t,\ln (S_0) +\mu t +\sigma x).
    \end{eqnarray*}
              
We also have
       \begin{eqnarray*}
\delta u(t,x)= \frac{\partial u}{\partial t}(t,x)+\frac{1}{2}   \frac{\partial^2 u}{\partial x^2} (t,x)  &=&
          e^{-rt}\left(-\frac{\partial U}{\partial t}+(A-r)U\right)(T-t,\ln (S_0) +\mu t +\sigma x)\\
          &=&
          e^{-rt}(A-r)\varphi (\ln (S_0) +\mu t +\sigma x)\ind{\{\ln (S_0) +\mu t +\sigma x\leq \tilde b(T-t)\}},
\end{eqnarray*}
where the last equality follows from regularity results (see, for instance, \cite{Jaillet}).
\end{remark}
We will need a more precise description of the operator $\cal D$, given by the following proposition (see Proposition 3.4 of \cite{DL2002}).
For convenience, we denote by $X$ a random variable with the same distribution as $X_1$,
 which is independent of the sequence $(X_n)_{n\geq 1}$.
\begin{prop}\label{prop3-2}
    Assume that (H4)  is satisfied and that $v$
    is a  function of class $C^3$ on $[0,T]\times \R$.
    For $0\leq t\leq T-h$ and $x\in\R$, define
    \[
   \tilde{{\cal D}}v(t,x)=2\int_0^{\sqrt{h}}d\xi\int_0^\xi dz\E\left[ X\left(\xi-X^2(\xi-z)\right)
                         {\partial^2 v\over \partial t\partial x}(t+\xi^2,x+zX)\right].
    \]
   We have
       \[
      {\cal D}v(t,x)=\tilde{{\cal D}}v(t,x)+2\int_0^{\sqrt{h}}d\xi\int_0^\xi dz
                   \E\left(X^2\delta v(t+\xi^2,x+zX)\right),
       \]
   with the notation $\delta v= {\partial v\over\partial t}+{1\over 2}{\partial^2 v\over\partial x^2}$, and
     \[
    \tilde{{\cal D}}v(t,x)=2\int_0^{\sqrt{h}}d\xi\int_0^\xi dz
              (\xi-z)\E\left[
              X^2\left(\xi-X^2\frac{(\xi-z)}{2}\right)
 {\partial^3 v\over \partial t\partial x^2}(t+\xi^2,x+zX)\right].
     \]
\end{prop}
\begin{remark}\label{rem2}\rm
Note that, if $\delta v(s,x+zX)=0$ for all $s\in[t,t+h]$ and $z\in[0,\sqrt{h}]$, we have $v(t,x)={\cal D}v(t,x)$.
\end{remark}
From the last equality in Proposition~\ref{prop3-2}, we derive the following estimates.
\begin{eqnarray*}
\left|\tilde{{\cal D}}v(t,x) \right|&\leq &
         2\int_0^{\sqrt{h}}\xi ^2d\xi\int_0^\xi dz
              \E\left[
              \left(X^2+\frac{X^4}{2}\right)
\left|{\partial^3 v\over \partial t\partial x^2}(t+\xi^2,x+zX)\right|\right]\\
   &\leq &\sqrt{h}\int_0^{\sqrt{h}} 2\xi d\xi 
              \E\left[\int_0^\xi dz
              \left(X^2+\frac{X^4}{2}\right)
\left|{\partial^3 v\over \partial t\partial x^2}(t+\xi^2,x+zX)\right|\right]\\
   &\leq &\sqrt{h}
        \int_t^{t+h} ds\E\left(\int dy \ind{\{|y-x|\leq \sqrt{h}|X|\}}\left(|X|+\frac{|X|^3}{2}\right)
           \left|{\partial^3 v\over \partial t\partial x^2}(s,y)\right|\right)\\
           &=&\sqrt{h}
        \int_t^{t+h} ds\int dy \E\left(\ind{\{|y-x|\leq \sqrt{h}|X|\}}\left(|X|+\frac{|X|^3}{2}\right)\right)
           \left|{\partial^3 v\over \partial t\partial x^2}(s,y)\right|
\end{eqnarray*}
We know from Proposition 3.2 of \cite{DL2002} (based on Berry-Esseen estimates) that, for every $k\in (1,3]$,
there exists a positive 
constant $C_{k}$ (which does not depend on $X$), 
such that, for all $y\in \R$, $n\geq 1$ and $j\in\{1,2,\ldots,n\}$,
\begin{eqnarray*}
\E\left(\left(|X|+\frac{|X|^3}{2}\right)
       \ind{\left\{\left|B^{(n)}_{jh}-y\right|\leq \sqrt{h}|X|\right\}}\right)
     &\leq &  \frac{C_{k}}{\sqrt{j}}{\E\left(|X|^3\right)\left(1+\E\;|X|^{3+k}\right)\over 1+|y|^k}.
\end{eqnarray*}
Hence, for $j=1,\ldots, n-1$,
\begin{eqnarray}
\E\left(\left|\tilde{{\cal D}}v(jh,B^{(n)}_{jh}) \right|\right)
            &\leq &  \frac{C_{k,X}}{\sqrt{j}}\sqrt{h}\int_{jh}^{jh+h} ds\int \frac{dy}{1+|y|^k} 
           \left|{\partial^3 v\over \partial t\partial x^2}(s,y)\right|\nonumber\\
           &\leq &C_{k,X}h\sqrt{2}\int_{jh}^{jh+h} \frac{ds}{\sqrt{s}}\int \frac{dy}{1+|y|^k} 
           \left|{\partial^3 v\over \partial t\partial x^2}(s,y)\right|,\label{Dtilde}
\end{eqnarray}
where, for the last inequality, we used the inequality $jh\geq (j+1)h/2$.

\section{Estimates for the second order time derivative}\label{Section-estimates}
In this section, we refine the regularity results that we used in \cite{DL2002}. We first establish
some elementary $L_1$-estimates. Then, we obtain a quadratic estimate for the second order time
derivative of the difference $\tilde U=U-\bar U$.
For the definition of the relevant weighted Sobolev spaces, we will use the notation
\[
\nu_j(dx)=\frac{dx}{(1+x^2)^{j/2}}, \quad j>1.
\]
\subsection{Some elementary $L_1$-estimates}
\begin{prop}\label{prop-L1}
      Assume that the function $\varphi$ is continuous and satisfies $\varphi\in L_1(\nu_j)$, 
       $\varphi'\in L_1(\nu_j)$
      and the second derivative $\varphi''$ is a Radon measure on $\R$, with $\int_\R \frac{|\varphi''(dz)|}{(1+z^2)^{j/2}}<\infty$.
      
      Let 
      \[
      u_\varphi(t,x)=e^{-rt}\E(\varphi(X^x_t)), \quad t\geq 0, \quad x\in \R.
      \]
      
      Then, for all $T>0$, there exists a constant $C_T>0$, such that
      \[
      \forall t\in (0,T], \quad \left|\left|\frac{\partial ^2u_\varphi}{\partial t^2}(t,.)
           \right|\right|_{L_1(\nu_j)}\leq \frac{C_T}{t}.
      \]
\end{prop}
We will easily deduce this proposition from the following lemma.
\begin{lemma}\label{lem-L1}
   If $\rho$ is a Radon measure on $\R$ and $q$ a nonnegative integrable function on $\R$, we have
   \[
      \left|\left|\rho*q\right|\right|_{L_1(\nu_j)}\leq 2^{j/2}\int_\R \frac{|\rho(dz)|}{(1+z^2)^{j/2}}\int_{-\infty}^\infty q(x) (1+x^2)^{j/2} dx.
   \]
   We also have, for any measurable function $f$ on $\R$,
   \[
   \forall y\in \R,\quad ||f(.-y)||_{L_1(\nu_j)}\leq 2^{j/2}(1+y^2)^{j/2}||f||_{L_1(\nu_j)}.
   \]
   \end{lemma}
\begin{pf}
 We have
 \begin{eqnarray*}
  \left|\left|\rho*q\right|\right|_{L_1(\nu_j)}&\leq &\int_{-\infty}^\infty  \frac{dx}{(1+x^2)^{j/2}}
         \int_\R| \rho(dz)| q(x-z)\\
        &=& \int_\R \frac{|\rho(dz)|}{ (1+z^2)^{j/2}}\int_{-\infty}^\infty q(x-z) \frac{(1+z^2)^{j/2}}{(1+x^2)^{j/2}}dx\\
        &=&\int_\R \frac{|\rho(dz)|}{ (1+z^2)^{j/2}}\int_{-\infty}^\infty q(x) \frac{(1+z^2)^{j/2}}{(1+(x+z)^2)^{j/2}}dx.
\end{eqnarray*}
Note that
\[
z^2\leq 2((x+z)^2+x^2),
\]
so that 
we deduce
 \begin{eqnarray*}
  \frac{1+z^2}{1+(x+z)^2}&\leq &\frac{1+2(x+z)^2+2x^2}{1+(x+z)^2}\\
             &\leq &2(1+x^2).
\end{eqnarray*}
Hence
\[
  \left|\left|\rho*q\right|\right|_{L_1(\nu_j)}\leq 2^{j/2}\int_\R \frac{|\rho(dz)|}{(1+z^2)^{j/2}}\int_{-\infty}^\infty q(x) (1+x^2)^{j/2} dx.
\]
Similarly, we have, for any measurable function $f$   and $y\in \R$, 
\begin{eqnarray*}
 \left|\left|f(.-y)\right|\right|_{L_1(\nu_j)}&=&\int |f(x-y)|\frac{dx}{(1+x^2)^{j/2}}\\
         &=&\int |f(x)|\frac{dx}{(1+(x+y)^2)^{j/2}}\\
         &=&\int |f(x)|\left(\frac{1+x^2}{1+(x+y)^2}\right)^{j/2}\frac{dx}{(1+x^2)^{j/2}}\\
         &\leq &
         \int |f(x)|\left(\frac{1+2(x+y)^2+2y^2}{1+(x+y)^2}\right)^{j/2}\frac{dx}{(1+x^2)^{j/2}}\\
         &\leq &
         2^{j/2}(1+y^2)^{j/2}\left|\left|f\right|\right|_{L_1(\nu_j)}.
\end{eqnarray*}

\end{pf}

\noindent {\bf Proof of Proposition~\ref{prop-L1}:}
We have
\begin{eqnarray*}
u_\varphi(t,x)&=&e^{-rt}\int_{-\infty}^\infty \varphi (x+y)\exp\left(-\frac{(y-\mu t)^2}{2\sigma^2 t}\right)\frac{dy}{\sigma\sqrt{2\pi t}}\\
    &=& e^{-rt}p_t*\varphi(x),
\end{eqnarray*}
with
\[
p_t(x)=\frac{1}{\sigma\sqrt{2\pi t}}\exp\left(-\frac{(x+\mu t)^2}{2\sigma^2 t}\right)=\frac{1}{\sigma\sqrt{ t}}n\left(\frac{x+\mu t}{\sigma\sqrt{t}}\right).
\]
Here, $n$ denotes the standard normal density function.

On the other hand, we know that $u_\varphi$ satisfies the equation 
\begin{equation}\label{eq-u}
 \frac{\partial u_\varphi}{\partial t}=(A-r)u_\varphi,
\end{equation}
so that
\begin{eqnarray*}
    \frac{\partial u_\varphi}{\partial t}(t,.)&=&e^{-rt}(A-r)p_t*\varphi\\
             &=&e^{-rt}p_t*[(A-r)\varphi].
\end{eqnarray*}
It follows from our assumptions that $(A-r)\varphi$ is a Radon measure satisfying 
\[
\int_\R|(A-r)\varphi(dz)|\frac{1}{(1+z^2)^{j/2}}<\infty.
\]
So that, using Lemma~\ref{lem-L1},
\begin{eqnarray*}
\left|\left|\frac{\partial u_\varphi}{\partial t}(t,.)\right|\right|_{L_1(\nu_j)}&\leq & 
            C_j \int_{-\infty}^\infty p_t(x) (1+|x|^j)dx\\
     &=&C_j \int_{-\infty}^\infty \frac{1}{\sigma\sqrt{ t}}n\left(\frac{x+\mu t}{\sigma\sqrt{t}}\right) (1+|x|^j)dx\\
     &=&C_j \int_{-\infty}^\infty n\left(y\right) (1+|y\sigma\sqrt{t}-\mu t|^j)dy\\
     &\leq & C_j\left(1+t^j\right).
\end{eqnarray*}
On the other hand, by differentiating \eqref{eq-u}, we have
\[
 \frac{\partial ^2 u_\varphi}{\partial t^2}=(A-r)\frac{\partial u_\varphi}{\partial t}
      = e^{-rt}\left((A-r)p_t\right)*(A-r)\varphi.
\]
Hence, using Lemma~\ref{lem-L1},  and the definition of $p_t$,
\begin{eqnarray*}
\left|\left|\frac{\partial ^2 u_\varphi}{\partial t^2}(t,.)\right|\right|_{L_1(\nu_j)}&\leq &C_j
         \int_{-\infty}^\infty |(A-r)p_t(x)| (1+|x|^j)dx\\
         &\leq &\frac{ C_j}{t}\left(1+t^j\right).
\end{eqnarray*}
\qed
\subsection{Quadratic estimates}
Recall the notation:
\[
U(t,x)= \sup_{\tau\in\T_{0,t}}\E\left(e^{-r\tau}\varphi(X^x_{\tau})\right),
\quad u_\varphi(t,x)=e^{-rt}\E(\varphi(X^x_t)), \quad t\geq 0, \quad x\in \R,
\]
with $\phi(x)=\left(K-e^x\right)^+$. We now introduce the difference $\tilde{U}=U-u_\varphi$ 
(which corresponds to the early exercise premium). We have the following
$L_2$-estimate for the second time derivative of $\tilde U=U-u_\varphi$.

\begin{thm}\label{thm-quadratic}
   Fix $T>0$ and $j>1$. There exists a constant $C>0$ such that, for all $\xi\in(0,T]$,
   \[
      \int_\xi^T(t-\xi)
      \left|\left|\frac{\partial ^2 \tilde U}{\partial t^2}(t,.)\right|\right|^2_{L_2(\nu_j)}dt\leq 
      C\left(1+|\ln \xi|^\beta\right),
         \quad\mbox{with } \beta=\left\{ \begin{array}{l}
                                                     3/2, \mbox{ if } d\leq r,\\
                                                     \\
                                                     1,  \mbox{ if } d>r.
                                                        \end{array}
                                                         \right.
   \]
\end{thm}
This estimate is closely related to Theorem~2.4 of \cite{DL2002}, a variant of results due to
Friedman and  Kinderlehrer (see \cite{Friedman1975}, Lemma 4.1,  and \cite{Kinderlehrer1980}, Chapter VIII). 
Note that by considering the difference $\tilde U=U-u_\varphi$, we are able to derive a logarithmic upper bound,
instead of a power of $\xi$, which would come up by considering $U$ (see Theorem~2.4 of \cite{DL2002}).
For the proof of Theorem~\ref{thm-quadratic}, we need some preliminary estimates on the derivatives 
$\frac{\partial ^2 \tilde U}{\partial x^2}$ and $\frac{\partial ^2 \tilde U}{\partial t\partial x}$.
\begin{lemma}\label{fact1}
   Fix $T>0$ and $j>1$. For any $\epsilon\in (0,1/4)$, there exists a constant $C>0$ such that, for all $t\in(0,T]$,
   \[
    \left|\left|\frac{\partial  \tilde U}{\partial x}(t,.)\right|\right|_{L_2(\nu_j)}\leq C \sqrt{t} \quad\mbox{and}\quad
      \left|\left|\frac{\partial ^2 \tilde U}{\partial x^2}(t,.)\right|\right|_{L_2(\nu_j)}\leq C t^\epsilon.
         \]
\end{lemma}
\begin{pf}
We know that $\tilde U$ solves the equation
\[
    -\frac{\partial  \tilde U}{\partial t}+(A-r)\tilde U=\tilde{h},
\]
with initial condition $\tilde U(0,.)=0$, where the function $\tilde{h}$ is given by
\[
\tilde{h}(t,x)=(A-r)\varphi(x)\ind{\{x\leq \tilde b(t)\}}, \quad t>0, \quad x\in \R.
\]
We have the following identity (which can be viewed as a form of the early exercise premium
formula).
\[
  \tilde U(t,.)=-\int_0^t e^{-r(t-s)} p_{t-s}*\tilde{h}(s,.)ds,
\]
where 
\[
p_t(x)=\frac{1}{\sigma\sqrt{2\pi t}}\exp\left( -\frac{(x+\mu t)^2}{2\sigma^2 t}\right)=\frac{1}{\sigma\sqrt{ t}}n\left(\frac{x+\mu t}{\sigma\sqrt{t}}\right),
\]
with $n$ denoting the standard normal density function.
It is straightforward to check that
\[
\frac{\partial  \tilde U}{\partial x}(t,.)=-\int_0^t e^{-r(t-s)} p_{t-s}*\frac{\partial \tilde{h}}{\partial x}(s,.)ds,
\]
and, with the notation $\delta_z$ for the Dirac measure at a point $z$,
\begin{eqnarray}
  \frac{\partial \tilde{h}}{\partial x}(t,x)&=&(A-r)\varphi'(x)\ind{\{x\leq \tilde b(t)\}}-(A-r)\varphi(x)\delta_{\tilde b(t)}(x)\nonumber\\
         &=& -\kappa(t,x)+\gamma(t)\delta_{\tilde b(t)}(x),\label{dhdx}
\end{eqnarray}
with $\kappa(t,x)=-(A-r)\varphi'(x)\ind{\{x\leq \tilde b(t)\}}$ and 
$\gamma(t)=-(A-r)\varphi(\tilde b(t))$. 
Note that $\kappa$ is a bounded function on $(0,\infty)\times \R$ and $\gamma$ is a continuous, nonnegative 
and bounded function on $(0,+\infty)$.
At this stage, it is clear that $||p_{t-s}*\frac{\partial \tilde{h}}{\partial x}(s,.)||_\infty\leq C/\sqrt{t-s}$, so that
\[
\left|\left|\frac{\partial  \tilde U}{\partial x}(t,.)\right|\right|_{L_2(\nu_j)}\leq C \sqrt{t}.
\]

On the other hand, we have
\begin{eqnarray*}
\left|\left|\frac{\partial ^2 \tilde U}{\partial x^2}(t,.)\right|\right|_{L_2(\nu_j)}
    &\leq & \int_0^t e^{-r(t-s)}
       \left|\left|p'_{t-s}*\kappa(s,.)\right|\right|_{L_2(\nu_j)}ds+\left|\left|\zeta(t,.)\right|\right|_{L_2(\nu_j)},
\end{eqnarray*}
with
\begin{eqnarray*}
   \zeta(t,.)&=& \int_0^t e^{-r(t-s)}\gamma(s) p'_{t-s}*\delta_{\tilde b(s)}ds\\
        &=& \int_0^t e^{-r(t-s)}\gamma(s) p'_{t-s}(.-\tilde b(s))ds.
\end{eqnarray*}
We have, using Lemma~\ref{lem-L1},
\begin{eqnarray*}
\left|\left|p'_{t-s}*\kappa(s,.)\right|\right|_{L_2(\nu_j)}&=&
           \left|\left|\int p'_{t-s}(y)\kappa(s,.-y)dy\right|\right|_{L_2(\nu_j)}\\
           &\leq&\int |p'_{t-s}(y)|\left|\left|\kappa(s,.-y)\right|\right|_{L_2(\nu_j)}dy\\
           &\leq &2^{j/4}\left|\left| \kappa(s,.)\right|\right|_{L_2(\nu_j)} \int |p'_{t-s}(y)|(1+y^2)^{j/4}dy.
\end{eqnarray*}
Note that, since $\kappa$ is bounded and $j>1$, $\sup_{s>0}\left|\left|\kappa(s,.)\right|\right|_{L_2(\nu_j)}<\infty$,
so that, for some constant $C>0$ (which may vary from line to line)
\begin{eqnarray*}
\left|\left|p'_{t-s}*\kappa(s,.)\right|\right|_{L_2(\nu_j)}&\leq&
       C\int |p'_{t-s}(y)|(1+y^2)^{j/4}dy\\
        &=&C\int \frac{1}{\sigma^2(t-s)}\left|n'\left(\frac{y+\mu(t-s)}{\sigma\sqrt{t-s}}\right)\right|(1+y^2)^{j/4}dy\\
        &=&C\int \frac{1}{\sigma\sqrt{t-s}}\left|n'\left(z\right)\right|(1+(-\mu(t-s)+\sigma\sqrt{t-s}z)^2)^{j/4}dz\\
        &\leq &\frac{C}{\sqrt{t-s}}\left( 1+ (t-s)^{j/2}\right).
\end{eqnarray*}
Hence, if $0<t<T$,
\[
\int_0^t e^{-r(t-s)}
       \left|\left|p'_{t-s}*\kappa(s,.)\right|\right|_{L_2(\nu_j)}ds\leq C\int_0^t \frac{ds}{\sqrt{t-s}}=2C\sqrt{t}.
\]
We now estimate $\left|\left|\zeta(t,.)\right|\right|_{L_2(\nu_j)}$.

We have, using the boundedness of $\gamma$, 
\begin{eqnarray*}
|\zeta(t,x)|&=&\left|\int_0^t e^{-r(t-s)}\gamma(s)p'_{t-s}(x-\tilde b(s))ds\right|\\
          &\leq& C\int_0^t \frac{1}{\sigma^2(t-s)}\left|n'\left(\frac{x-\tilde b(s)+\mu(t-s)}{\sigma\sqrt{t-s}}\right)\right|ds
\end{eqnarray*}
Recall that $n'(x)=-xn(x)$. Therefore
\begin{eqnarray*}
|\zeta(t,x)|&\leq& C\int_0^t \frac{|x-\tilde b(s)+\mu(t-s)|}{(t-s)^{3/2}}n\left(\frac{x-\tilde b(s)+\mu(t-s)}{\sigma\sqrt{t-s}}\right)ds\\
       &\leq &C\int_0^t\frac{ds}{\sqrt{t-s}}+C\int_0^t \frac{|x-\tilde b(s)|}{(t-s)^{3/2}}n\left(\frac{x-\tilde b(s)+\mu(t-s)}{\sigma\sqrt{t-s}}\right)ds.
\end{eqnarray*}
Note that 
\begin{align*}
n(x_1+x_2)=n(x_1)\exp\left(-\frac{x_2^2}{2}-x_1x_2\right)&\leq n(x_1)\exp(-x_1x_2)\\
                                                     &\leq n(x_1)\exp\left(\frac{x_1^2}{4}+x_2^2\right)
     =n(x_1/\sqrt{2})e^{x_2^2}.
\end{align*}
Hence, for $t\in (0,T)$,
\begin{eqnarray*}
|\zeta(t,x)|
       &\leq &C\int_0^t\frac{ds}{\sqrt{t-s}}+
           C_T\int_0^t \frac{|x-\tilde b(s)|}{(t-s)^{3/2}}n\left(\frac{x-\tilde b(s)}{\sqrt{2}\sigma\sqrt{t-s}}\right)ds.
\end{eqnarray*}
Note that, for all $\alpha>0$, there exists $C_\alpha>0$, such that, for all $y\in \R$, $n(y/\sqrt{2})\leq C_\alpha/|y|^{2\alpha}$.
Hence, for $t\in (0,T)$,
\begin{eqnarray*}
|\zeta(t,x)|
       &\leq &C\sqrt{t}+C_\alpha\int_0^t\frac{|x-\tilde b(s)|}{(t-s)^{3/2}}\frac{(t-s)^\alpha}{|x-\tilde b(s)|^{2\alpha}}ds\\
       &=& C\sqrt{t}+C_\alpha \int_0^t\frac{(t-s)^{\alpha-\frac{3}{2}}}{|x-\tilde b(s)|^{2\alpha-1}}ds\\
       &=&C\sqrt{t}+C_\alpha t^{\alpha-\frac{1}{2}}\int_0^1\frac{1}{(1-u)^{\frac{3}{2}-\alpha}|x-\tilde b(tu)|^{2\alpha-1}}du.
\end{eqnarray*}   
Now, take $\alpha=\frac{1}{2}+\epsilon$ (with $0<\epsilon<1/4$) and put $\beta(t,x)=|x-\tilde b(t)|^{1-2\alpha} =|x-\tilde b(t)|^{-2\epsilon}$.
We get
\begin{eqnarray*}
\left|\left|\zeta(t,.)\right|\right|_{L_2(\nu_j)}
       &\leq 
      &C\sqrt{t}+Ct^{\epsilon}\int_0^1\frac{1}{(1-u)^{1-\epsilon}}\left|\left|\beta(tu,.)\right|\right|_{L_2(\nu_j)}du.
\end{eqnarray*} 
Using Lemma~\ref{lem-L1}, we have
\[
\left|\left|\beta(tu,.)\right|\right|_{L_2(\nu_j)}^2\leq 2^{j/2}\left(1+\tilde b(tu)^2\right)^{j/2}\int \frac{1}{|x|^{4\epsilon}}\frac{dx}{(1+x^2)^{j/2}}.
\]
Since $\epsilon <1/4$, the integral on the righthand side is finite, and the lemma easily follows.
\end{pf}

We now turn to the study of $\frac{\partial ^2 \tilde U}{\partial t\partial x}$. 
Recall that $\partial U/\partial t$ solves
the parabolic equation $-\partial v/\partial t +(A-r)v=0$ in the set $\{(t,x)\;|\; t>0, x>\tilde b(t)\}$. 
Since the exercise 
boundary is differentiable and $\partial U/\partial t$ is continuous and vanishes on the exercise boundary, it follows
that $\frac{\partial^2 U}{\partial t\partial x}$ is continuous ``up to the boundary", i.e. on the set
$\{(t,x)\;|\; t>0, x\geq \tilde b(t)\}$ (see \cite{Friedman1975}, Lemma 4.5). We first show that $\frac{\partial^2 U}{\partial t\partial x}$
is nonnegative along the exercise boundary.
\begin{lemma}\label{crossderiv}
 We have, for any $t>0$,
 \[
 \frac{\partial^2 U}{\partial t\partial x}(t,\tilde b(t))\geq 0.
 \]
\end{lemma}
\begin{pf}
  We have, for all $t>0$, due to the {\em smooth fit} property,
  \[
  \frac{\partial U}{\partial x}(t,\tilde b(t))=\varphi'(\tilde b(t)),
  \]
  so that, by differentiating with respect to $t$,
  \[
  \frac{\partial^2 U}{\partial t\partial x}(t,\tilde b(t))+\frac{\partial^2 U}{\partial x^2}(t,\tilde b(t))\tilde b'(t)=\varphi''(\tilde b(t))\tilde b'(t)
  \]
  and
  \[
  \frac{\partial^2 U}{\partial t\partial x}(t,\tilde b(t))=-\left(\frac{\partial^2 U}{\partial x^2}(t,\tilde b(t))-\varphi''(\tilde b(t))\right)\tilde b'(t).
  \]
  Observe that, for each $t>0$, the function $x\mapsto U(t,x)-\varphi(x)$ is $C^2$ on the interval
  $[\tilde b(t),\infty)$ and has a minimum at $\tilde b(t)$. Therefore, its second derivative must be nonnegative at
  this point. Since $\tilde b'(t)\leq 0$, the lemma is proved.
\end{pf}
\begin{lemma}\label{fact2}
   Fix $T>0$ and $j>1$. There exists a constant $C>0$ such that, for all $t_1\in(0,T\wedge 1]$,
   \[
      \int_{t_1}^T\left|\left|\frac{\partial ^2 \tilde U}{\partial t\partial x}(t,.)\right|\right|_{L_2(\nu_j)}^2dt\leq C\ln(1/t_1).
         \]
\end{lemma}
For the proof of Lemma~\ref{fact2}, we will need the bilinear form associated with the operator $A-r$.

We first introduce the relevant weighted Sobolev spaces. For $j>1$,
let $H_j=L^2(\R,\nu_j)$ and
$V_j=\{f\in H_j \;|\; f'\in H_j\}$. The inner product on $H_j$
will be denoted by $(\cdot,\cdot)_j$ and the associated norm
by $|\cdot|_j$. The natural norm on $V_j$ will be denoted 
by $||\cdot||_j$. Thus, we have
\[
|f|^2_j=\int_{-\infty}^{+\infty}f^2(x)\frac{dx}{(1+x^2)^{j/2}},
\]
and $||f||^2_j=|f|^2_j+|f'|^2_j$.

Recall that    the partial differential operator $A$ is defined by
\[
A=\frac{\sigma^2}{2} {\partial ^2 \over \partial x^2}
          +\mu {\partial \over \partial x}.
\]
We associate with the operator $A-r$ a bilinear functional on $V_j$,
defined by
\begin{eqnarray*}
a_j(f,g)&=&\frac{\sigma^2}{2} \int_{-\infty}^{\infty}f'(x)g'(x)\frac{dx}{(1+x^2)^{j/2}}
      -\frac{j\sigma^2}{2}\int_{-\infty}^{\infty}f'(x)g(x)\frac{x}{(1+x^2)^{(j/2)+1}}dx\\
     && -\mu\int_{-\infty}^{\infty}f'(x)g(x)\frac{dx}{(1+x^2)^{j/2}}
      +r\int_{-\infty}^{\infty}f(x)g(x)\frac{dx}{(1+x^2)^{j/2}},
\end{eqnarray*}
so that, if $f'\in V_j$, 
\[
a_j(f,g)=-((A-r)f,g)_j.
\]
It will be convenient to write $a_j(f,g)$ as
$
a_j(f,g)= \tilde{a}_j(f,g)+\bar{a}_j(f,g),
$
with
\begin{equation}\label{eq-atilde}
\tilde{a}_j(f,g)=\frac{\sigma^2}{2}\left[(f',g')_j+(f,g)_j\right]\quad\mbox{and}
      \quad \bar{a}_j(f,g)=a_j(f,g)-\tilde{a}_j(f,g).
\end{equation}
With these notations, it is easy to check that $|\bar{a}_j(f,g)|\leq C ||f||_j |g|_j$ 
and $|\bar{a}_j(f,g)|\leq C ||g||_j |f|_j$, for some constant $C$
which does not depend on $f$ nor $g$. 

\medskip

\noindent {\bf Proof of Lemma~\ref{fact2}:}
In order to rule out regularity issues, we introduce  a $C^\infty$, nonnegative function $\rho$
on $\R\times \R$, with $\int \rho(t,x)dtdx=1$ and $\mbox{supp }\rho \subset [-1,0]\times [-1,+1]$ and set, for any positive
integer $m$, 
$\rho_m(t,x)= m^2\rho(mt,mx)$.

Now, let $W=\frac{\partial\tilde U}{\partial x}$, $W_m=W*\rho_m$, and $h_m=\tilde{h}*\rho_m$. For each $m>0$,
the functions $W_m$, $h_m$ are $C^\infty$ with bounded derivatives and we have
\[
-\frac{\partial W_m}{\partial t}+(A-r)W_m=\frac{\partial h_m}{\partial x}.
\]
Multiply by $\partial W_m/\partial t$ and integrate with respect to $\nu_j$ to get,
for any fixed $t>0$,
\[
-\left(\frac{\partial W_m}{\partial t}(t,.), \frac{\partial W_m}{\partial t}(t,.)\right)_j
-a_j\left(W_m(t,.), \frac{\partial W_m}{\partial t}(t,.)\right)=\int \frac{\partial h_m}{\partial x}(t,x)\frac{\partial W_m}{\partial t}(t,x)\nu_j(dx).
\]
Note that 
\begin{eqnarray*}
a_j\left(W_m(t,.), \frac{\partial W_m}{\partial t}(t,.)\right)&=&
    \tilde{a}_j\left(W_m(t,.), \frac{\partial W_m}{\partial t}(t,.)\right)+\bar{a}_j\left(W_m(t,.), \frac{\partial W_m}{\partial t}(t,.)\right)\\
    &=&
    \frac{1}{2}\frac{d}{dt}\left(\tilde a_j\left(W_m(t,.), W_m(t,.)\right)\right)+\bar{a}_j\left(W_m(t,.), \frac{\partial W_m}{\partial t}(t,.)\right).
\end{eqnarray*}
By integrating with respect to time, we get, if $0<t_1<T$,
\begin{align*}
-\int_{t_1}^T\left|\frac{\partial W_m}{\partial t}(t,.)\right|_j^2dt+\frac{1}{2}\left(\tilde a_j\left(W_m(t_1,.), W_m(t_1,.)\right)
       -\tilde a_j\left(W_m(T,.), W_m(T,.)\right)\right)&=\\
     \int_{t_1}^T\bar{a}_j\left(W_m(t,.), \frac{\partial W_m}{\partial t}(t,.)\right)dt
     +\int_{t_1}^T\left(\frac{\partial h_m}{\partial x}(t,.),\frac{\partial W_m}{\partial t}(t,.)\right)_j&dt.
\end{align*}
Hence
\begin{eqnarray*}
\int_{t_1}^T\left|\frac{\partial W_m}{\partial t}(t,.)\right|_j^2dt&\leq &
      \frac{1}{2}\tilde a_j\left(W_m(t_1,.), W_m(t_1,.)\right)-
          \int_{t_1}^T\bar{a}_j\left(W_m(t,.), \frac{\partial W_m}{\partial t}(t,.)\right)dt\\
          &&
     -\int_{t_1}^T\left(\frac{\partial h_m}{\partial x}(t,.),\frac{\partial W_m}{\partial t}(t,.)\right)_jdt\\
     &\leq &\frac{1}{2}\tilde a_j\left(W_m(t_1,.), W_m(t_1,.)\right)+
    C \int_{t_1}^T\left|\left| W_m(t,.)\right|\right|_j\left| \frac{\partial W_m}{\partial t}(t,.)\right|_jdt \\
  &&  -\int_{t_1}^T\left(\frac{\partial h_m}{\partial x}(t,.),\frac{\partial W_m}{\partial t}(t,.)\right)_jdt.
\end{eqnarray*}
Using the inequality 
\[
2\left|\left| W_m(t,.)\right|\right|_j\left| \frac{\partial W_m}{\partial t}(t,.)\right|_j\leq 
      \epsilon \left| \frac{\partial W_m}{\partial t}(t,.)\right|_j^2  +\frac{1}{\epsilon}\left|\left| W_m(t,.)\right|\right|_j^2,
 \]
 we get
 \begin{eqnarray*}
\frac{1}{2}\int_{t_1}^T\left|\frac{\partial W_m}{\partial t}(t,.)\right|_j^2dt&\leq &
       \frac{1}{2}\tilde a_j\left(W_m(t_1,.), W_m(t_1,.)\right)+
      C\int_{t_1}^T\left|\left| W_m(t,.)\right|\right|_j^2dt   \\
      &&
     - \int_{t_1}^T\left(\frac{\partial h_m}{\partial x}(t,.),\frac{\partial W_m}{\partial t}(t,.)\right)_jdt.
\end{eqnarray*}
We have
\begin{eqnarray*}
\tilde a_j\left(W_m(t_1,.), W_m(t_1,.)\right)&\leq &C\left|\left| W_m(t_1,.)\right|\right|_j^2
\end{eqnarray*}
and, using Lemma~\ref{lem-L1}, 
\begin{eqnarray*}
\left|\left| W_m(t_1,.)\right|\right|_j&\leq &\int \rho_m(t_1-t,y)\left|\left| W(t,.-y)\right|\right|_j dtdy\\
&\leq&\int \rho_m(t_1-t,y)2^{j/4}(1+y^2)^{j/4}\left|\left| W(t,.)\right|\right|_jdtdy.
\end{eqnarray*}
Using Lemma~\ref{fact1},  we have (for $\epsilon\in (0,1/4)$) 
$\left|\left| W(t,.)\right|\right|_j\leq Ct^\epsilon$. Hence
\begin{eqnarray*}
\tilde{a}_j\left(W_m(t_1,.), W_m(t_1,.)\right)&\leq &C
    \int \rho_m(t_1-t,y)2^{j/2}(1+y^2)^{j/2}t^{2\epsilon} dtdy\\
    &=&C
    \int \rho_m(t,y)2^{j/2}(1+y^2)^{j/2}|t_1-t|^{2\epsilon} dtdy   \\
    &=&
    C\int \rho(t,y)2^{j/2}\left(1+\frac{y^2}{m^2}\right)^{j/2}\left|t_1-\frac{t}{m}\right|^{2\epsilon} dtdy\\
    &\leq &C(1+t_1^{2\epsilon}).
\end{eqnarray*}
   
   We also have
\begin{eqnarray*}
\int_{t_1}^T\left|\left| W_m(t,.)\right|\right|_j^2dt &=&
        \int_{t_1}^T\left|\left| \int dsdy \rho_m(t-s, y)W(s,.-y)\right|\right|_j^2dt\\
&\leq &
       \int_{t_1}^T\int dsdy \rho_m(t-s, y)\left|\left| W(s,.-y)\right|\right|_j^2dt\\
&\leq &
          \int_{t_1}^Tdt \int dsdy \rho_m(t-s, y)
               2^{j/2}(1+y^2)^{j/2}\left|\left| W(s,.)\right|\right|_j^2\\
          &\leq &
          \int_{t_1}^{T+\frac{1}{m}}ds\left|\left| W(s,.)\right|\right|_j^2\int dtdy
            \rho_m(t-s, y)2^{j/2}(1+y^2)^{j/2}.
\end{eqnarray*}
Since $\int_0^{T +1}\left|\left| W(s,.)\right|\right|_j^2ds<\infty$, we deduce  that
 \begin{eqnarray*}
\frac{1}{2}\int_{t_1}^T\left|\frac{\partial W_m}{\partial t}(t,.)\right|_j^2dt&\leq &
C\left(1+t_1^{2\epsilon}\right)
-\int_{t_1}^T\left(\frac{\partial h_m}{\partial x}(t,.),\frac{\partial W_m}{\partial t}(t,.)\right)_jdt.
\end{eqnarray*}
It follows from the proof of Lemma~\ref{fact1} (see \eqref{dhdx}) that
\[
\frac{\partial h_m}{\partial x}(t,x)=-\kappa_m(t,x)+\gamma_m(t,x),
\]
where $\kappa_m=\kappa *\rho_m$, and $\kappa$ is a bounded function, and
\[
\gamma_m(t,x)=\int \rho_m(t-\tau,x-\tilde b(\tau))\gamma(\tau)d\tau.
\]
Hence
\begin{eqnarray*}
- \int_{t_1}^T\left(\frac{\partial h_m}{\partial x}(t,.),\frac{\partial W_m}{\partial t}(t,.)\right)_jdt&\leq &
     C\int_{t_1}^T\left|\frac{\partial W_m}{\partial t}(t,.)\right|_jdt -
       \int_{t_1}^T\left( \gamma_m(t,.),\frac{\partial W_m}{\partial t}(t,.)\right)_jdt\\
       &\leq&\frac{1}{4}\int_{t_1}^T\left|\frac{\partial W_m}{\partial t}(t,.)\right|^2_jdt+C^2T -
       \int_{t_1}^T\left( \gamma_m(t,.),\frac{\partial W_m}{\partial t}(t,.)\right)_jdt,
\end{eqnarray*}
where we have used $C\left|\frac{\partial W_m}{\partial t}(t,.)\right|_j\leq \frac{1}{4}\left|\frac{\partial W_m}{\partial t}(t,.)\right|_j^2+C^2$.
Therefore
 \begin{eqnarray}
\frac{1}{4}\int_{t_1}^T\left|\frac{\partial W_m}{\partial t}(t,.)\right|_j^2dt&\leq &
C\left(1+t_1^{2\epsilon}\right)- \int_{t_1}^T\left(\gamma_m(t,.),\frac{\partial W_m}{\partial t}(t,.)\right)_j dt. \label{eq-Jm}
\end{eqnarray}
Note that 
\[
\frac{\partial W_m}{\partial t}=\frac{\partial^2 U_m}{\partial t\partial x}-
                \frac{\partial^2 u_m}{\partial t\partial x},
\]
where
\[
U_m=U*\rho_m\quad \mbox{and}\quad u_m=u_\varphi*\rho_m,
\]
so that
\[
- \int_{t_1}^T\left(\gamma_m(t,.),\frac{\partial W_m}{\partial t}(t,.)\right)_jdt
=J^{(1)}_m+J^{(2)}_m,
\]
with
\[
J^{(1)}_m=-\int_{t_1}^T\left(\gamma_m(t,.),\frac{\partial ^2U_m}{\partial t\partial x}(t,.)
\right)_jdt
\quad \mbox{and}\quad
J^{(2)}_m=\int_{t_1}^T\left(\gamma_m(t,.),\frac{\partial ^2 u_m}{\partial t\partial x}(t,.)
 \right)_jdt.
 \]
 We have
 \begin{eqnarray*}
J^{(1)}_m&=&
-\int_{t_1}^Tdt\int \frac{dx}{(1+x^2)^{j/2}}\gamma_m(t,x)
                           \frac{\partial ^2U_m}{\partial t\partial x}(t,x)\\
                           &=&-\int_{t_1}^Tdt\int \frac{dx}{(1+x^2)^{j/2}}
                           \int d\tau \gamma(\tau)\rho_m(t-\tau, x-\tilde b(\tau))
                           \frac{\partial ^2U_m}{\partial t\partial x}(t,x)\\
                           &=&
                           -\int d\tau \int ds \int dy \ind{\{t_1<\tau+s<T\}}\gamma(\tau)
                           \rho_m(s, y)
                           \frac{\partial ^2U_m}{\partial t\partial x}(\tau+s,\tilde b(\tau)+y)
                           \frac{1}{(1+(y+\tilde b(\tau))^2)^{j/2}}\\
                           &=&-\int_{t_1}^{T+\frac{1}{m}}\gamma(\tau)\eta_m(\tau)d\tau, 
\end{eqnarray*}
where
\begin{eqnarray*}
\eta_m(\tau)&=&\int_{t_1-\tau}^{T-\tau}ds\int \frac{dy}{(1+(y+\tilde b(\tau))^2)^{j/2}}
        \rho_m(s,y)\frac{\partial ^2U_m}{\partial t\partial x}(\tau+s,\tilde b(\tau)+y)\\
        &=&
      \int_{t_1-\tau}^{T-\tau}ds\int \frac{\rho_m(s,y)dy}{(1+(y+\tilde b(\tau))^2)^{j/2}}
        \int \int ds' dy'
         \rho_m(s',y')\frac{\partial ^2U}{\partial t\partial x}(\tau+s-s',\tilde b(\tau)+y-y').  
\end{eqnarray*}
Note that $\frac{\partial ^2U}{\partial t\partial x}=0$ on the open set 
$S=\{(t,x)\;|\; t>0, x<\tilde b(t)\}$ (which is the interior set of the stopping region),
so that
\begin{eqnarray*}
\eta_m(\tau)&=&
\int_{t_1-\tau}^{T-\tau}ds\int \frac{\rho_m(s,y)dy}{(1+(y+\tilde b(\tau))^2)^{j/2}}
        \int \int ds' dy'
         \rho_m(s',y')\bar W(\tau,s-s',y-y'),
 \end{eqnarray*}
 where
 \[
 \bar W(\tau,\theta,z)=\frac{\partial ^2U}{\partial t\partial x}(\tau+\theta,\tilde b(\tau)+z)
                    \ind{\{\tilde b(\tau)+z\geq \tilde b(\tau+\theta)\}}.
 \]
 Since $\frac{\partial ^2U}{\partial t\partial x}(\tau,\tilde b(\tau))\geq 0$, for $\tau>0$, we have
 \begin{eqnarray*}
\eta_m(\tau)&\geq&\int_{t_1-\tau}^{T-\tau}ds\int\int\int
      \frac{\rho_m(s,y)dy\rho_m(s',y')ds'dy'}{\left(1+(y+\tilde b(\tau))^2\right)^{j/2}}
     D(\tau,s-s',y-y'),
 \end{eqnarray*}
 where
 \[
 D(\tau,\theta,z)=\left(
     \frac{\partial ^2U}{\partial t\partial x}(\tau+\theta,\tilde b(\tau)+z)-\frac{\partial ^2U}{\partial t\partial x}(\tau,\tilde b(\tau))
               \right)\ind{\{\tilde b(\tau)+z\geq \tilde b(\tau+\theta)\}}.
 \]
 Hence (since $\gamma\geq 0$)
\begin{eqnarray*}
J^{(1)}_m&\leq&-\int_{t_1}^{T+\frac{1}{m}} \gamma(\tau)\epsilon_m(\tau)d\tau,
\end{eqnarray*}
 with
 \[
 \epsilon_m(\tau)=\int_{t_1-\tau}^{T-\tau}ds\int\int\int
      \frac{\rho_m(s,y)dy\rho_m(s',y')ds'dy'}{(1+(y+\tilde b(\tau))^2)^{j/2}}
     D(\tau,s-s',y-y').
 \]
 We have
 \[
 | \epsilon_m(\tau)|\leq \int_{t_1-\tau}^{T-\tau}D_m(\tau)ds=(T-t_1)D_m(\tau),
 \]
 where
 \[
 D_m(\tau)=\sup_{|\theta|\leq 1/m, |z|\leq 2/m }\left(\left|D(\tau,\theta,z)
                       \right|\ind{\{\tilde b(\tau)+z\geq \tilde b(\tau+\theta)\}}\right).
 \]
 Due to the continuity properties of $\frac{\partial^2 U}{\partial t\partial x}$, 
 as $m\to \infty$,
 the function $D_m$ converges to $0$, uniformly on the interval $[t_1, T+1]$.
 Therefore, we have
 \begin{equation}\label{eq-Jm1}
 \limsup_{m\to \infty}J^{(1)}_m\leq 0.
 \end{equation}
 We now examine $J^{(2)}_m$. We have, using the boundedness of $\gamma$,
 \begin{eqnarray*}
|J^{(2)}_m|&\leq &\int_{t_1}^Tdt\int \frac{dx}{(1+x^2)^{j/2}}|\gamma_m(t,x)|
                           \left|\frac{\partial ^2 u_m}{\partial t\partial x}(t,x)\right|\\
                           &\leq &
                           C\int_{t_1}^Tdt\int \frac{dx}{(1+x^2)^{j/2}}
                           \int d\tau \rho_m(t-\tau, x-\tilde b(\tau))
                          \left| \frac{\partial ^2 u_m}{\partial t\partial x}(t,x)\right|.
\end{eqnarray*}
Note that, since $\varphi$ is Lipschitz, we have 
$\left|\left|  \frac{\partial ^2 u_\varphi}{\partial t\partial x}(t,.)\right|\right|_{\infty}\leq \frac{C}{t}$ and,
since $\mbox{supp }\rho \subset [-1,0]\times [-1,+1]$,
\begin{align*}
\left|\left|  \frac{\partial ^2 u_m}{\partial t\partial x}(t,.)\right|\right|_{\infty}\leq
\int\int d\tau dy \rho_m(\tau,y) 
     \left|\left|  \frac{\partial ^2 u_\varphi}{\partial t\partial x}(t-\tau,.)\right|\right|_{\infty}\leq \frac{C}{t}.
\end{align*}
Hence
\begin{eqnarray}\label{eq-Jm2}
|J^{(2)}_m|&\leq &C\int_{t_1}^T\frac{dt}{t}\int dx
                           \int d\tau \rho_m(t-\tau, x-\tilde b(\tau))=C\ln\frac{T}{t_1}.
 \end{eqnarray}
 It follows from \eqref{eq-Jm}, \eqref{eq-Jm1} and \eqref{eq-Jm2} that
 \begin{eqnarray*}
\limsup_{m\to \infty}\int_{t_1}^T\left|\frac{\partial W_m}{\partial t}(t,.)\right|_j^2dt&\leq &
      C\left(1+t_1^{2\epsilon}+\ln\frac{T}{t_1}\right),
\end{eqnarray*}
which proves the lemma.
 \qed
 \subsection{Proof of Theorem~\ref{thm-quadratic}}
 For the proof of Theorem~\ref{thm-quadratic}, we will work on the equation satisfied
 by $\partial \tilde U/\partial t$. Let
 \[
 V=\frac{\partial \tilde U}{\partial t}.
 \]
 We have
 \[
 -\frac{\partial V}{\partial t}+(A-r)V=\frac{\partial \tilde{h}}{\partial t},
 \]
 where
 \[
\tilde{h}(t,x)=(A-r)\varphi(x)\ind{\{x\leq \tilde b(t)\}}, \quad t>0, \quad x\in \R.
\]
 The following lemma will clarify the computation of  the  derivative $\partial \tilde{h}/\partial t$
 in the sense of distributions.
 \begin{lemma}\label{lem-dhdt}
 Define the function $I$ on $(0,+\infty)\times \R$ by
 \[
 I(t,x)=\ind{\{x\leq \tilde b(t)\}}, \quad t>0, \quad x\in \R.
 \]
 The distribution $\partial I/\partial t$ applied to a compactly supported $C^\infty$
 function $\rho$ on $(0,+\infty)\times \R$ is given by
 \[
 \langle \frac{\partial I}{\partial t}, \rho\rangle=\int  \tilde b'(t)\rho(t,\tilde b(t))dt.
 \]
 This can be written (less precisely): $\frac{\partial I}{\partial t}(t,.)=\tilde b'(t)
            \delta_{\tilde b(t)}$.
 \end{lemma}
 \begin{pf}
 We have
 \begin{eqnarray*}
  \langle \frac{\partial I}{\partial t}, \rho\rangle&=&
           -\langle I, \frac{\partial \rho}{\partial t} \rangle   \\
           &=&-\int dt\int dx I(t,x) \frac{\partial \rho}{\partial t}(t,x)
\end{eqnarray*}
Let $J$ be the range of $\tilde b$. We have $J=(\tilde b(\infty),\tilde b(0))$.
Note that, if $x\leq \tilde b(\infty)$, $I(t,x)=1$ for all $t>0$ and if $x\geq \tilde b(0)$
$I(t,x)=0$ for all $t>0$, so that, in both cases, 
$\int I(t,x) \frac{\partial \rho}{\partial t}(t,x)dt=0$. Therefore
\begin{eqnarray*}
  \langle \frac{\partial I}{\partial t}, \rho\rangle&=&
           -\int_J dx \int dt \ind{\{x\leq \tilde b(t)\}} \frac{\partial \rho}{\partial t}(t,x)\\
           &=&
           -\int_J dx \int dt \ind{\{t\leq \tilde b^{-1}(x)\}} \frac{\partial \rho}{\partial t}(t,x)\\
           &=&
           -\int_J dx  \rho(\tilde b^{-1}(x),x)\\
           &=&
           \int \tilde b'(t)\rho(t,\tilde b(t))dt.
\end{eqnarray*}
Here, we have used the fact that $\tilde b$ is strictly decreasing (which is proved in 
\cite{Villeneuve1999}), but we can also approximate $\tilde b$ by the strictly
decreasing functions $\tilde b_\epsilon(t)=-\epsilon t+\tilde b(t)$ to derive the formula.
In fact, we only need $\tilde b$ to be $C^1$: indeed, we can replace 
$\tilde b(t)$ by $\tilde b_\mu(t)=-\mu t+\tilde b(t)$ and choose $\mu$ so that $\tilde b_\mu$
is strictly increasing in a neighborhood of the time projection of the support of $\rho$.
 \end{pf}
 
 We now proceed with the proof of Theorem~\ref{thm-quadratic}. 
 As in the proof of Lemma~\ref{fact2}, we introduce a regularizing sequence $\rho_m$,
 and set
 \[
 V_m=V*\rho_m \quad \mbox{and}\quad \chi_m=\frac{\partial \tilde{h}}{\partial t}*\rho_m,
 \]
 so that
 \[
  -\frac{\partial V_m}{\partial t}+(A-r)V_m=\chi_m
   \]
   Note that the functions $V_m$, $\chi_m$ are $C^\infty$, with bounded derivatives on any subset
   $[t_1,T]\times\R$, with $0<t_1<T$. This is due to the fact that $V$ is bounded on such subsets.
   For any fixed $t>0$, multiply by $\partial V_m/\partial t$ and integrate with respect
   to $\nu_j$ to get
   \[
   -\left| \frac{\partial V_m}{\partial t}(t,.)\right|_j^2
        -a_j\left(V_m(t,.), \frac{\partial V_m}{\partial t}(t,.)\right)=\int \chi_m(t,x)
            \frac{\partial V_m}{\partial t}(t,x)\nu_j(dx).
   \]
   We have
 \begin{eqnarray*}
a_j\left(V_m(t,.), \frac{\partial V_m}{\partial t}(t,.)\right)    &=&
    \frac{1}{2}\frac{d}{dt}\left(\tilde a_j\left(V_m(t,.), V_m(t,.)\right)\right)+
    \bar{a}_j\left(V_m(t,.), \frac{\partial V_m}{\partial t}(t,.)\right).
\end{eqnarray*}
By integrating with respect to time, we get, if $0<t_1<T$,
\begin{eqnarray*}
\lefteqn{
-\int_{t_1}^{T}\left|\frac{\partial V_m}{\partial t}(t,.)\right|_j^2dt+
\frac{1}{2}\left[\tilde a_j\left(V_m(t_1,.), V_m(t_1,.)\right)
       -\tilde a_j\left(V_m(T,.), V_m(T,.)\right)\right]=}\\
     && \;\;\;\int_{t_1}^{T}\bar{a}_j\left(V_m(t,.), \frac{\partial V_m}{\partial t}(t,.)\right)dt
     +\int_{t_1}^{T}\left(\chi_m(t,.),\frac{\partial V_m}{\partial t}(t,.)\right)_jdt.
\end{eqnarray*}
Hence
\begin{eqnarray*}
\int_{t_1}^{T}\left|\frac{\partial V_m}{\partial t}(t,.)\right|_j^2dt&\leq&
    C\left|\left|V_m(t_1,.)\right|\right|_j^2
    -\int_{t_1}^{T}\bar{a}_j\left(V_m(t,.), \frac{\partial V_m}{\partial t}(t,.)\right)dt\\
    &&
     -\int_{t_1}^{T}\left(\chi_m(t,.),\frac{\partial V_m}{\partial t}(t,.)\right)_jdt\\
     &\leq&
     C\left(\left|\left|V_m(t_1,.)\right|\right|_j^2+
          \int_{t_1}^{T}\left|\left|V_m(t,.)\right|\right|_j
          \left| \frac{\partial V_m}{\partial t}(t,.)\right|_jdt\right)+J_m(t_1,T),
\end{eqnarray*}
with
\[
J_m(t_1,T)=-\int_{t_1}^{T}\left(\chi_m(t,.),\frac{\partial V_m}{\partial t}(t,.)\right)_jdt.
\]
Using the inequality 
\[
2\left|\left|V_m(t,.)\right|\right|_j
          \left| \frac{\partial V_m}{\partial t}(t,.)\right|_j
          \leq
          \epsilon \left| \frac{\partial V_m}{\partial t}(t,.)\right|_j^2+\frac{1}{\epsilon}
          \left|\left|V_m(t,.)\right|\right|_j^2,
\]
 we derive
 \begin{eqnarray}
\frac{1}{2}\int_{t_1}^{T}\left|\frac{\partial V_m}{\partial t}(t,.)\right|_j^2dt&\leq& 
C\left(\left|\left|V_m(t_1,.)\right|\right|_j^2+
          \int_{t_1}^{T}\left|\left|V_m(t,.)\right|\right|_j^2dt\right) +J_m(t_1,T)\label{final}.
\end{eqnarray}   
We now study  $J_m(t_1,T)$.
Note that, for any fixed $t>0$,
\begin{eqnarray*}
\left(\chi_m(t,.),\frac{\partial V_m}{\partial t}(t,.)\right)_j&=&
   \int \nu_j(dx) \frac{\partial V_m}{\partial t}(t,x)
       \frac{\partial \tilde{h}}{\partial t}*\rho_m(t,x)
\end{eqnarray*}
  We have
  \[
   \frac{\partial \tilde{h}}{\partial t}(t,.)=(A-r)\varphi \frac{\partial I}{\partial t}(t,.),
  \]
  so that,  using Lemma~\ref{lem-dhdt}, and the notation
  $\gamma(t)=-(A-r)\varphi(\tilde b(t))$
 \[
  \frac{\partial \tilde{h}}{\partial t}*\rho_m(t,x)=-\int d\tau \rho_m(t-\tau,x-\tilde b(\tau))
       \tilde b'(\tau)\gamma(\tau),
 \] 
Recall that $\gamma(\tau)\geq 0$.
 Hence
\begin{eqnarray*}
\left(\chi_m(t,.),\frac{\partial V_m}{\partial t}(t,.)\right)_j&=&
     -\int \nu_j(dx) \frac{\partial V_m}{\partial t}(t,x)
     \int d\tau \rho_m(t-\tau,x-\tilde b(\tau))
       \tilde b'(\tau) \gamma(\tau)\\
     &=& - \int d\tau\int \frac{dx}{(1+x^2)^{j/2}}
       \frac{\partial V_m}{\partial t}(t,x)\rho_m(t-\tau,x-\tilde b(\tau))
       \tilde b'(\tau) \gamma(\tau)\\
       &=&-\int d\tau\int \frac{dy}{(1+(y+\tilde b(\tau))^2)^{j/2}}
       \frac{\partial V_m}{\partial t}(t,y+\tilde b(\tau))\rho_m(t-\tau,y)
       \tilde b'(\tau)\gamma(\tau).
\end{eqnarray*} 
Going back to $J_m(t_1, T)$, we have
\begin{eqnarray*}
J_m(t_1,T)&=&-\int_{t_1}^T dt\int d\tau \int dy 
     \frac{\partial V_m}{\partial t}(t,y+\tilde b(\tau))
        \rho_m(t-\tau,y)\bar\gamma_j(\tau,y),
\end{eqnarray*}
with
\[
\bar\gamma_j(\tau,y)=-\frac{1}{(1+(y+\tilde b(\tau))^2)^{j/2}}\tilde b'(\tau) \gamma(\tau).
\]
Note that $\bar\gamma_j(\tau, y)\geq 0$.
We have
\begin{eqnarray*}
J_m(t_1,T)&=&-\int d\tau\int dt \int dy \ind{\{t_1<t<T\}}
     \frac{\partial V_m}{\partial t}(t,y+\tilde b(\tau))
        \rho_m(t-\tau,y)\bar\gamma_j(\tau,y)\\
        &=&-\int d\tau\int ds \int dy \ind{\{t_1<\tau +s<T\}}
     \frac{\partial V_m}{\partial t}(\tau+s,y+\tilde b(\tau))
        \rho_m(s,y)\bar\gamma_j(\tau,y).
\end{eqnarray*}
Observe that
\[
\frac{d}{d\tau}\left(V_m(\tau+s,y+\tilde b(\tau))\right)=
     \frac{\partial V_m}{\partial t}(\tau+s,y+\tilde b(\tau))
     +\frac{\partial V_m}{\partial x}(\tau+s,y+\tilde b(\tau))\tilde b'(\tau),
\]
so that
\[
\frac{\partial V_m}{\partial t}(\tau+s,y+\tilde b(\tau))=
   \frac{d}{d\tau}\left(V_m(\tau+s,y+\tilde b(\tau))\right)-
   \frac{\partial V_m}{\partial x}(\tau+s,y+\tilde b(\tau))\tilde b'(\tau).
\]
Hence
\[
J_m(t_1,T)=\hat J_m(t_1,T)+\bar J_m(t_1,T),
\]
with
\[
\hat J_m(t_1,T)=
     -\int d\tau\int ds \int dy \ind{\{t_1<\tau +s<T\}}
     \frac{d}{d\tau}\left(V_m(\tau+s,y+\tilde b(\tau))\right)
        \rho_m(s,y)\bar\gamma_j(\tau,y)
\]
and
\[
\bar J_m(t_1,T)=
    +\int d\tau\int ds \int dy \ind{\{t_1<\tau +s<T\}}
     \frac{\partial V_m}{\partial x}(\tau+s,y+\tilde b(\tau))\tilde b'(\tau)
        \rho_m(s,y)\bar\gamma_j(\tau,y).
\]
We have, using integration by parts,
\begin{eqnarray*}
\hat J_m(t_1,T)&=&-\int ds\int dy \rho_m(s,y)\left(\int_{t_1-s}^{T-s}
              \frac{d}{d\tau}\left(V_m(\tau+s,y+\tilde b(\tau))\right)\bar\gamma_j(\tau,y)d\tau
              \right)\\
              &=&
            -\int ds\int dy \rho_m(s,y)
            V_m(T,y+\tilde b(T-s))\bar\gamma_j(T-s,y)
            \\
            &&
            +\int ds\int dy \rho_m(s,y)V_m(t_1,y+\tilde b(t_1-s))\bar\gamma_j(t_1-s,y)\\
            &&
            +\int ds\int dy \rho_m(s,y)\int_{t_1-s}^{T-s}V_m(s+\tau,y+\tilde b(\tau))
            \frac{\partial \bar\gamma_j}{\partial \tau}(\tau,y)d\tau.
\end{eqnarray*}
Note that, due to the continuity of $V(=\partial \tilde U/\partial t)$ on $(0,\infty)\times\R$,
the sequence $V_m$ converges uniformly to $V$ on compact sets.
We also have the continuity of $\bar\gamma_j$ and $\partial \bar\gamma_j/\partial \tau$
(due to the fact that $\tilde b$ is $C^2$). We easily deduce thereof that
\begin{eqnarray}
\lim_{m\to\infty}\hat J_m(t_1,T)=-V(T,\tilde b(T))\bar\gamma_j(T,0)+
V(t_1,\tilde b(t_1))\bar\gamma_j(t_1,0)+
\int_{t_1}^{T}V(\tau,\tilde b(\tau))
            \frac{\partial \bar\gamma_j}{\partial \tau}(\tau,0)d\tau,\label{Jhat}
\end{eqnarray}
and the convergence is uniform with respect to $t_1$, as long as $t_1$ remains in a compact 
set of the form $[\xi,T]$, where $0<\xi<T$.
For $\bar J_m(t_1, T)$, we have
\begin{eqnarray*}
  \bar J_m(t_1, T)&=&\bar J^{(1)}_m(t_1, T)+\bar J^{(2)}_m(t_1, T),
\end{eqnarray*}
with
\[
\bar J^{(1)}_m(t_1, T)=+\int d\tau\int ds \int dy \ind{\{t_1<\tau +s<T\}}
     \frac{\partial^2 U_m}{\partial t\partial x}(\tau+s,y+\tilde b(\tau))\tilde b'(\tau)
        \rho_m(s,y)\bar\gamma_j(\tau,y)
\]
and
\[
\bar J^{(2)}_m(t_1, T)=-\int d\tau\int ds \int dy \ind{\{t_1<\tau +s<T\}}
     \frac{\partial^2 u_m}{\partial t\partial x}(\tau+s,y+\tilde b(\tau))\tilde b'(\tau)
        \rho_m(s,y)\bar\gamma_j(\tau,y).
\]
We deal with $\bar J^{(1)}_m(t_1, T)$ in the same way as for the proof of \eqref{eq-Jm1}.
Using the fact that $\tilde b'(\tau)\bar\gamma_j(\tau,y)\leq 0$, we have $\bar J^{(1)}_m(t_1, T)\leq\tilde J^{(1)}_m(t_1, T)$,
with
\begin{eqnarray*}
\tilde J^{(1)}_m(t_1, T)&=&\int d\tau\int ds \int dy \ind{\{t_1<\tau +s<T\}}\int\int ds'dy'\rho_m(s',y')D(\tau,s-s',y-y')
     \tilde b'(\tau)
        \rho_m(s,y)\bar\gamma_j(\tau,y),
\end{eqnarray*}
where
\[
D(\tau,\theta,z)= \left(\frac{\partial^2 U}{\partial t\partial x}(\tau+\theta,\tilde b(\tau)+z)-
        \frac{\partial^2 U}{\partial t\partial x}(\tau,\tilde b(\tau))\right)\ind{\{\tilde b(\tau)+z\geq \tilde b(\tau+\theta)\}}.
\]
Due to the continuity properties of $\frac{\partial^2 U}{\partial t\partial x}$, we have
 \[
 \lim_{m\to \infty}\tilde J^{(1)}_m(t_1, T)= 0,
 \]
 and the convergence is uniform with respect to $t_1$, as long as $t_1$ remains in $[\xi,T]$.
 On the other hand, due to the continuity of
 $ \frac{\partial^2 u_\varphi}{\partial t\partial x}$, we have
 \[
 \lim_{m\to \infty}\bar J^{(2)}_m(t_1, T)
    =-\int_{t_1}^Td\tau
     \frac{\partial^2 u_\varphi}{\partial t\partial x}(\tau,\tilde b(\tau))\tilde b'(\tau)
        \bar\gamma_j(\tau,0),
 \]
 uniformly with respect to $t_1\in[\xi,T]$. At this stage, we can state that $J_m(t_1,T)\leq \tilde J_m(t_1,T)$, with
 $\tilde J_m(t_1,T)=\hat J_m(t_1,T)+ \tilde J^{(1)}_m(t_1, T)+\bar J^{(2)}_m(t_1, T)$, and
\[
\lim_{m\to \infty}\sup_{t_1\in [\xi,T]} \left|\tilde J_m(t_1, T)-\tilde J(t_1,T)\right|=0,
\]
 where
 \begin{eqnarray*}
\tilde J(t_1, T)&=&
   -V(T,\tilde b(T))\bar\gamma_j(T,0)+
V(t_1,\tilde b(t_1))\bar\gamma_j(t_1,0)+
\int_{t_1}^{T}V(\tau,\tilde b(\tau))
            \frac{\partial \bar\gamma_j}{\partial \tau}(\tau,0)d\tau\\
            &&
            -\int_{t_1}^Td\tau
     \frac{\partial^2 u}{\partial t\partial x}(\tau,\tilde b(\tau))\tilde b'(\tau)
        \bar\gamma_j(\tau,0).
\end{eqnarray*}
Since $\partial U/\partial t$ vanishes along the exercise boundary, we have
$V(t,\tilde b(t))=-\frac{\partial u_\varphi}{\partial t}(t,\tilde b(t))$, so that
\begin{eqnarray*}
\lefteqn{
-V(T,\tilde b(T))\bar\gamma_j(T,0)+
V(t_1,\tilde b(t_1))\bar\gamma_j(t_1,0)+
\int_{t_1}^{T}V(\tau,\tilde b(\tau))
            \frac{\partial \bar\gamma_j}{\partial \tau}(\tau,0)d\tau=}\\
          &&  \frac{\partial u_\varphi}{\partial t}(T,\tilde b(T))\bar\gamma_j(T,0)-
\frac{\partial u_\varphi}{\partial t}(t_1, \tilde b(t_1))\bar\gamma_j(t_1,0)-
\int_{t_1}^{T}\frac{\partial u_\varphi}{\partial t}(\tau,\tilde b(\tau))
            \frac{\partial \bar\gamma_j}{\partial \tau}(\tau,0)d\tau\\
            &=&\int_{t_1}^{T}\frac{d}{d\tau}\left(
            \frac{\partial u_\varphi}{\partial t}(\tau,\tilde b(\tau))\right)
             \bar\gamma_j(\tau,0)d\tau,
\end{eqnarray*}
so that
\begin{eqnarray*}
\tilde J(t_1, T)&=&
    \int_{t_1}^{T}
    \left[\frac{d}{d\tau}\left(
            \frac{\partial u_\varphi}{\partial t}(\tau,\tilde b(\tau))\right)
            -\frac{\partial^2 u_\varphi}{\partial t\partial x}(\tau,\tilde b(\tau))\tilde b'(\tau)\right]
             \bar\gamma_j(\tau,0)d\tau\\
             &=&
             \int_{t_1}^{T}\frac{\partial^2 u_\varphi}{\partial t^2}(\tau,\tilde b(\tau))\bar\gamma_j(\tau,0)
             d\tau.
\end{eqnarray*}
We now go back to \eqref{final} and integrate with respect to $t_1$ to derive
 \begin{eqnarray*}
\frac{1}{2}\int_\xi^Tdt_1\int_{t_1}^{T}\left|\frac{\partial^2 \tilde U_m}{\partial t^2}(t,.)\right|_j^2dt&\leq& 
          C\left(\int_\xi^T\left|\left|V_m(t_1,.)\right|\right|_j^2dt_1+
          \int_\xi^T\left(\int_{t_1}^{T}\left|\left|V_m(t,.)\right|\right|_j^2dt \right)dt_1\right)\\
        && + 
          \int_{\xi}^Tdt_1\tilde J_m(t_1,T).
\end{eqnarray*}   
Hence
\begin{eqnarray*}
\frac{1}{2}\int_{\xi}^{T}(t-\xi)\left|\frac{\partial^2 \tilde U_m}{\partial t^2}(t,.)\right|_j^2dt&\leq& 
          C\int_\xi^T\left|\left|V_m(t,.)\right|\right|_j^2dt
          + 
        \int_\xi^Tdt_1\tilde J_m(t_1,T).
\end{eqnarray*} 
Note that 
\begin{eqnarray*}
\lim_{m\to \infty}\int_\xi^T\left|\left|V_m(t,.)\right|\right|_j^2dt&=&\int_\xi^T\left|\left|V(t,.)\right|\right|_j^2dt\\
   &\leq & C\left(1+\ln \frac{T}{\xi}\right),
\end{eqnarray*}
where the last inequality follows from Lemma~\ref{fact2}.
Moreover,
\begin{eqnarray*}
\lim_{m\to \infty} \int_\xi^Tdt_1\tilde J_m(t_1,T)&=& \int_\xi^Tdt_1\tilde J(t_1,T)\\
    &=&\int_{\xi}^{T}(t-\xi)\frac{\partial^2 u_\varphi}{\partial t^2}(t,\tilde b(t))\bar\gamma_j(t,0)
             dt.
\end{eqnarray*}

Hence
\begin{eqnarray*}
\frac{1}{2}\int_{\xi}^{T}(t-\xi)\left|\frac{\partial^2 \tilde U}{\partial t^2}(t,.)\right|_j^2dt&\leq& 
C\left(1+\ln \frac{T}{\xi}\right)+
             \int_{\xi}^{T}(t-\xi)\frac{\partial^2 u_\varphi}{\partial t^2}(t,\tilde b(t))\bar\gamma_j(t,0)
            dt.
\end{eqnarray*} 
Theorem~\ref{thm-quadratic} now follows from the following lemma, which relies on the asymptotic behavior
of the exercice boundary near maturity (see \cite{Barles}, \cite{LambertonVilleneuve}).
\begin{lemma}
We have
\begin{eqnarray*}
\int_{\xi}^{T}(t-\xi)\left|\frac{\partial^2 u_\varphi}{\partial t^2}(t,\tilde b(t))\right||\tilde b'(t)|dt
&\leq &C\left(1+|\ln \xi|^\beta\right),
         \quad\mbox{with } \beta=\left\{ \begin{array}{l}
                                                     3/2, \mbox{ if } d\leq r,\\
                                                     \\
                                                     1,  \mbox{ if } d>r.
                                                        \end{array}
                                                         \right.
\end{eqnarray*}
\end{lemma}
\begin{pf}
We first note that, since $\varphi $ is bounded and Lipschitz continuous, we have
\[
\left|\left|\frac{\partial^2 u_\varphi}{\partial t^2}(t,.)\right|\right|_\infty
\leq
\frac{C}{t^{3/2}}.
\]
This can be seen by arguing, as in the proof of Proposition~\ref{prop-L1}, that
$u_\varphi(t,.)=e^{-rt}p_t*\varphi$, so that
and $\frac{\partial^2 u_\varphi}{\partial t^2}(t,.)=(A-r)^2u_\varphi$. In order to estimate
the $x$-derivatives of $u_\varphi$ up to the order $4$, we may differentiate $p_t$ three times
and use the boundedness of $\varphi'$.

We then have

\begin{eqnarray*}
\int_{\xi}^{T}(t-\xi)\left|\frac{\partial^2 u_\varphi}{\partial t^2}(t,\tilde b(t))\right||\tilde b'(t)|dt&\leq&
   C\int_{\xi}^{T}\frac{t-\xi}{t^{3/2}}|\tilde b'(t)|dt\\
   &\leq &
   C\int_{\xi}^{T}\frac{1}{\sqrt{t}}|\tilde b'(t)|dt.
\end{eqnarray*}
Now, since $\tilde b'(t)\leq 0$, we have
\begin{eqnarray*}
  \int_{\xi}^{T}\frac{1}{\sqrt{t}}|\tilde b'(t)|dt&=&-\int_{\xi}^{T}\frac{1}{\sqrt{t}}\tilde b'(t)dt\\
  &=& -\left(\frac{\tilde b(T)-\tilde b(0)}{\sqrt{T}}-
          \frac{\tilde b(\xi)-\tilde b(0)}{\sqrt{\xi}}\right)-\frac{1}{2}\int_{\xi}^{T}
          \frac{1}{t^{3/2}}(\tilde b(t)-\tilde b(0))dt\\
          &\leq&\frac{\tilde b(0)-\tilde b(T)}{\sqrt{T}}+
          \frac{1}{2}\int_{\xi}^{T}\frac{1}{t^{3/2}}(\tilde b(0)-\tilde b(t))dt
\end{eqnarray*}
If $d\leq r$, we have $\tilde b(0)-\tilde b(t)\leq C\sqrt{t|\ln t|}$ for $t$ close to $0$,
so that 
\[
\int_{\xi}^{T}\frac{1}{t^{3/2}}(\tilde b(0)-\tilde b(t))dt\leq C
   \int_{\xi}^{T}\frac{1}{t}\sqrt{|\ln t|}dt\leq C(1+|\ln \xi|^{3/2}).
   \]
   If $d>r$, we have $\tilde b(0)-\tilde b(t)\leq C\sqrt{t}$ for $t$ close to $0$,
so that 
\[
\int_{\xi}^{T}\frac{1}{t^{3/2}}(\tilde b(0)-\tilde b(t))dt\leq C
   \int_{\xi}^{T}\frac{1}{t}dt= C\ln (T/\xi). 
   \]
\end{pf}

\section{Upper bound for $P^{(n)}_0-P_0$} \label{UB}
In order to derive an upper bound for $P^{(n)}_0-P_0$, we relate this quantity to the modified
value function $u$ using \eqref{*} as follows:
\begin{eqnarray*}
   P^{(n)}_0-P_0&=&\sup_{\tau\in{\cal T}^{(n)}_{0,T}}
      \E\left(
        e^{-r\tau}g(\mu_0\tau + B^{(n)}_\tau)\right) -u(0,0)\\
        &\leq &\sup_{\tau\in{\cal T}^{(n)}_{0,T}}
      \E\left(
        u(\tau , B^{(n)}_\tau)-u(0,0)\right) \\
        &=&\sup_{\tau\in{\cal T}^{(n)}_{0,T}}
      \E\left(
        \sum_{j=1}^{\tau/h}{\cal D}u({(j-1)h},B^{(n)}_{{(j-1)h}})\right).
  \end{eqnarray*}
We observe that ${\cal D}u\leq \tilde{{\cal D}}u$, and recall from \cite{DL2002} (Lemma 4.1) that $\sup_{0\leq j\leq n-1}\E\left|{\cal D}u({jh},B^{(n)}_{{jh}})\right|\leq Ch$, so that
\begin{eqnarray}
   P^{(n)}_0-P_0&\leq &
         \E\left(
        \sum_{j=1}^{n-2}\left|\tilde{{\cal D}}u({jh},B^{(n)}_{{jh}})\right|\right)+O(h)\nonumber\\
        &\leq&C_{k,X}h\sqrt{2}\int_{h}^{T-h} \frac{ds}{\sqrt{s}}\int \frac{dy}{1+|y|^k} 
           \left|{\partial^3 u\over \partial t\partial x^2}(s,y)\right|+O(h).\label{int*}
\end{eqnarray}

Here, we have a regularity problem, since $u$ is not $C^3$. This problem can be fixed as follows.
By convolution, one can approximate $u$ by a sequence $u_m$ which is smooth, uniformly bounded and satisfies $\delta u_m\leq 0$,
and  ${\cal D}u_m\leq \tilde{{\cal D}}u_m$. We need the following variant of Lemma~\ref{lem-L1}.
\begin{lemma}\label{lem-L1*}
   If $\rho$ is a Radon measure on $(0,T)\times\R$ and $q$ a nonnegative integrable function on $(0,T)\times\R$, 
   with $q(t,x)=0$ for $t\notin (0,a)$, where $a$ satisfies $0<a<h$, we have
   \[
      \int_{h}^{T-h} \frac{ds}{\sqrt{s}}\int \frac{\left|\rho*q(s,y)\right|}{(1+|y|^2)^{k/2}} 
           dy
      \leq 2^{k/2}\int_{h-a}^{T-h}\int_\R \frac{|\rho(dt, dz)|}{\sqrt{t}(1+z^2)^{k/2}}\int_0^ads\int_{-\infty}^\infty q(s,x) (1+x^2)^{k/2} dx.
   \]
      \end{lemma}
      \begin{pf}
We have
\begin{align*}
 \int_{h}^{T-h} \frac{ds}{\sqrt{s}}\int \frac{dy}{(1+|y|^2)^{k/2}} 
          & \left|\rho*q(s,y)\right|\leq \\
           & \int_{h}^{T-h} \frac{ds}{\sqrt{s}}\int \frac{dy}{(1+|y|^2)^{k/2}} 
          \int\int \left|\rho(dt,dz)\right| q(s-t,y-z)\\
          &=
          \int\int \left|\rho(dt,dz)\right|\int_{h}^{T-h} \frac{ds}{\sqrt{s}}\int \frac{dy}{(1+|y|^2)^{k/2}} q(s-t,y-z)\\
          &\leq \int_{h-a}^{T-h}\int \left|\rho(dt,dz)\right|\int_0^a\frac{d\theta}{\sqrt{t+\theta}}\int
           \frac{dy}{(1+|y|^2)^{k/2}} q(\theta,y-z)\\
           &\leq\int_{h-a}^{T-h}\int\frac{ \left|\rho(dt,dz)\right|}{\sqrt{t}}\int_0^a d\theta
          \int \frac{dy}{(1+|y|^2)^{k/2}} q(\theta,y-z)\\
           &=\int_{h-a}^{T-h}\int\frac{ \left|\rho(dt,dz)\right|}{\sqrt{t}(1+z^2)^{k/2}}\int_0^a d\theta
          \int  \frac{(1+z^2)^{k/2}dx}{(1+|x+z|^2)^{k/2}} q(\theta,x)\\
           &\le
           2^{k/2}\int_{h-a}^{T-h}\int\frac{ \left|\rho(dt,dz)\right|}{\sqrt{t}(1+z^2)^{k/2}}\int_0^a d\theta
          \int dx (1+x^2)^{k/2} q(\theta,x),
\end{align*}
where the last inequality follows from $ \frac{1+z^2}{1+(x+z)^2}\leq 2(1+x^2)$.
      \end{pf}
      
      Using Lemma~\ref{lem-L1*} and the fact that $\partial^3 u/(\partial t\partial x^2)$ is a Radon measure
      (see  \eqref{eq-deriv} and the comment below), we derive the correct version of \eqref{int*},
      namely
 \begin{eqnarray}
   P^{(n)}_0-P_0&\leq &
         C_{k,X}h\sqrt{2}\int_{h}^{T-h} \frac{1}{\sqrt{s}}\int \frac{1}{1+|y|^k} 
           \left|{\partial^3 u\over \partial t\partial x^2}(ds,dy)\right|+O(h).\label{int**}
\end{eqnarray}  

If we introduce the function $\tilde u:=u-\bar u$, we have, using the fact that $\delta \bar u=0$,
\begin{eqnarray*}
\int_{h}^{T-h} \frac{1}{\sqrt{s}}\int \frac{1}{1+|y|^k} 
           \left|{\partial^3 u\over \partial t\partial x^2}(ds,dy)\right|&\leq&
           \int_{h}^{T-h} \frac{1}{\sqrt{s}}\int \frac{1}{1+|y|^k} 
           \left|{\partial^3 \tilde u\over \partial t\partial x^2}(ds,dy)\right|\\
           &&
           +2\int_{h}^{T-h} \frac{ds}{\sqrt{s}}\int \frac{dy}{1+|y|^k} 
           \left|{\partial^2 \bar u\over \partial t^2}(s,y)\right|\\
           &\leq &\int_{h}^{T-h} \frac{1}{\sqrt{s}}\int \frac{1}{1+|y|^k} 
           \left|{\partial^3 \tilde u\over \partial t\partial x^2}(ds,dy)\right|\\
           &&
            +2C_T\int_{h}^{T-h} \frac{ds}{\sqrt{s}(T-s)}\\
            &\leq &\int_{h}^{T-h} \frac{1}{\sqrt{s}}\int \frac{1}{1+|y|^k} 
           \left|{\partial^3 \tilde u\over \partial t\partial x^2}(ds,dy)\right|+C_T|\ln h|,
\end{eqnarray*}
 where we have used Proposition~\ref{prop-L1}. 
 
 We now need to estimate $\int_{h}^{T-h} \frac{1}{\sqrt{s}}\int \frac{1}{1+|y|^k} 
           \left|{\partial^3 \tilde u\over \partial t\partial x^2}(ds,dy)\right|$. Recall from 
 Remark~\ref{rem1} that 
 \begin{eqnarray*}
\frac{\partial \tilde u}{\partial t}(t,x)+\frac{1}{2}   \frac{\partial^2 \tilde u}{\partial x^2} (t,x)            &=&
          \zeta(t,x)\ind{\{x\leq \hat b(t)\}},
\end{eqnarray*}
where
$\zeta(t,x)=e^{-rt}(A-r)\varphi (\ln (S_0) +\mu t +\sigma x)$ and 
$\hat b(t)=\left(\tilde b(T-t)-\mu t-\ln(S_0)\right)/\sigma$.
By differentiating wit respect to $t$, we derive the following expression
\begin{eqnarray}
\frac{\partial^2 \tilde u}{\partial t^2}(t,x)+\frac{1}{2}   \frac{\partial^3 \tilde u}{\partial t\partial x^2} (t,x)            &=&
         \frac{\partial \zeta}{\partial t}(t,x)\ind{\{x\leq \hat b(t)\}}+\zeta(t,\hat b(t))\hat b'(t)\delta_{\hat b(t)},\label{eq-deriv}
\end{eqnarray}
where we have used Lemma~\ref{lem-dhdt}.
Note that
\[
\sup_{x\leq \hat b(t)}\left|\frac{\partial \zeta}{\partial t}(t,x)\right|<\infty\mbox{ and }
\sup_{0<t<T}\left|\zeta(t,\hat b(t))\right|<\infty.
\]
Moreover, $|\hat b'(t)|\leq |\tilde b'(T-t)|+|\mu/\sigma|$, so that
\begin{eqnarray*}
\int_0^T\frac{dt}{\sqrt{t}}|\hat b'(t)|&\leq &
            \int_0^{T/2}\frac{dt}{\sqrt{t}}|\tilde b'(T-t)|+\int_{T/2}^T\frac{dt}{\sqrt{t}}|\tilde b'(T-t)|+2|\mu/\sigma| \sqrt{T}\\
            &\leq &\sup_{0\leq t\leq T/2}|\tilde b'(T-t)|\int_0^{T/2}\frac{dt}{\sqrt{t}}+
               +\sqrt{\frac{2}{T}}\int_{T/2}^T|\tilde b'(T-t)|dt+2|\mu/\sigma| \sqrt{T}\\
               &=&\sup_{0\leq t\leq T/2}|\tilde b'(T-t)|\int_0^{T/2}\frac{dt}{\sqrt{t}}+\sqrt{\frac{2}{T}}\left(\tilde b(0)-\tilde b(T/2)\right)+
               2|\mu/\sigma| \sqrt{T}   <\infty.
\end{eqnarray*}
Therefore, the righthand side of \eqref{eq-deriv} is a Radon measure and since, due to
Theorem~\ref{thm-quadratic}, 
$\partial^2 \tilde u/\partial t^2$ is locally integrable, it follows that
$\partial^3 \tilde u/\partial t\partial x^2$ is a Radon measure.

Moreover, we have
\begin{eqnarray*}
\int_{h}^{T-h} \frac{1}{\sqrt{s}}\int \frac{1}{1+|y|^k} 
           \left|{\partial^3 \tilde u\over \partial t\partial x^2}(ds,dy)\right|&\leq &
            C_T+2\int_{h}^{T-h} \frac{ds}{\sqrt{s}}\int \frac{1}{1+|y|^k} 
           \left|{\partial^2 \tilde u\over \partial t^2}(s,y)\right|.
\end{eqnarray*}
Now, using the Cauchy-Schwarz inequality and Theorem~\ref{thm-quadratic}, we have
\begin{align*}
\int_{h}^{T-h} \frac{ds}{\sqrt{s}}\int \frac{dy}{1+|y|^k} 
           \left|{\partial^2 \tilde u\over \partial t^2}(s,y)\right|&\leq\\
           C\left(\int_{h}^{T-h}\frac{ds}{s(T-s-\frac{h}{2})}\right)^{1/2}&\left(\int_{h}^{T-h}\!\!ds (T-s-\frac{h}{2})\left|{\partial^2 \tilde u\over \partial t^2}(s,.)\right|_k^2  \right)^{1/2}\\
           &\leq
           C\sqrt{|\ln h|}\left(\int_{h}^{T-\frac{h}{2}}\!\!ds (T-s-\frac{h}{2})\left|{\partial^2 \tilde u\over \partial t^2}(s,.)\right|_k^2  \right)^{1/2}\\
           &= C\sqrt{|\ln h|}\left(\int_{h/2}^{T-h}\!\!dt (t-\frac{h}{2})\left|{\partial^2 \tilde u\over \partial t^2}(T-t,.)\right|_k^2  \right)^{1/2}\\
           &\leq
           C\sqrt{|\ln h|^{1+\beta}},
\end{align*}
with $\beta=1$ if $d>r$ and $\beta=3/2$ if $d\leq r$. The last inequality follows from Theorem~\ref{thm-quadratic} and Lemma~\ref{fact2},
and the connection between the derivatives of the functions $\tilde{U}$ and $\tilde{u}$ (see Remark~\ref{rem1}; we also use the classical bounds
$||\partial U/\partial t(t,.)||_\infty+||\partial^2 U/\partial x^2(t,.)||_\infty\leq C/\sqrt{t}$). 
We conclude that
\[
P^{(n)}_0-P_0\leq C\frac{(\ln n)^\alpha}{n},
\]
with $\alpha=1$ if $d>r$ and $\alpha=5/4$ if $d\leq r$.
\section{Lower bound for $P^{(n)}_0-P_0$}\label{LB}
For the derivation of the lower bound, we use the stopping time introduced in 
\cite{DL2002} (see the proof of Theorem 5.6). Namely
\[
\tau=\tau_1\ind{\{\tau_1<T-h\}}+T\ind{\{\tau_1=T-h\}},
\]
where 
\[
\tau_1=\inf\left\{ t\in [0,T-h]\;\;|\;\; t/h   \in \N
       \quad\mbox{and}\quad d(B^{(n)}_t, I_{t+h})\leq 
                \sqrt{h}||X||_{\infty}+|\mu_0|h\right\}.
\]
Here, $I_t=\{x\in \R\;|\; u(t,x)=g(t,x+\mu_0 t)\}$. Note that, if $t<T$, $I_t=(-\infty, \hat b(t)]$.

The modification of $\tau_1$ into $\tau$ is motivated by the unboundedness of
$\partial u/\partial t$ near $T$.
We have, due to the definition of $\tau_1$, 
\begin{eqnarray*}
P^{(n)}-P&\geq &\E\left(e^{-r\tau}g(\mu_0 \tau+B^{(n)}_\tau)-u(0,0)\right)\\
         &=&\E\left(e^{-r\tau}g(\mu_0 \tau+B^{(n)}_\tau)-u(\tau,B^{(n)}_\tau)+u(\tau,B^{(n)}_\tau)-u(0,0)\right).
\end{eqnarray*}
We have
\begin{eqnarray*}
\E\left(u(\tau,B^{(n)}_\tau)-u(0,0)\right)&=&\E\left(\sum_{j=1}^{\tau/h}{\cal D}u((j-1)h,B^{(n)}_{(j-1)h})\right)\\
       &=&\E\left(\sum_{j=1}^{(\tau/h)\wedge(n-2)}{\cal D}u(jh-h,B^{(n)}_{jh-h})\right)\\
       &&+
               \E\left(\ind{\{\tau =T\}}\sum_{j=n-1}^n{\cal D}u(jh-h,B^{(n)}_{jh-h})\right)\\
               &=&\E\left(\sum_{j=1}^{(\tau/h)\wedge(n-2)}{\cal D}u(jh-h,B^{(n)}_{jh-h})\right)+O(h),
\end{eqnarray*}
where the last equality follows from Lemma~4.1 of \cite{DL2002}. Now, if $j<(\tau/h)\wedge(n-2)$,
we have $d(B^{(n)}_{jh}, I_{jh+h})> \sqrt{h}||X||_{\infty}+|\mu_0|h$, so that
\[
B^{(n)}_{jh}>\hat{b}(jh+h)+\sqrt{h}||X||_{\infty}+|\mu_0|h.
\]
We then have, for $s\in [jh,jh+h]$ and $z\in[0,\sqrt{h}]$
\begin{eqnarray*}
B^{(n)}_{jh}+zX&>&\hat{b}(jh+h)+zX+\sqrt{h}||X||_{\infty}+|\mu_0|h\\
                     &\geq &\hat{b}(jh+h)+|\mu_0|h\\
                   &=&\hat{b}(s)+\hat{b}(jh+h)+\mu_0(jh+h)-(\hat b(s)+\mu_0s)-\mu_0(jh+h-s)+|\mu_0|h\\
                   &\geq &\hat{b}(s),
\end{eqnarray*}
the last inequality coming from the fact that $t\mapsto \hat{b}(t)+\mu_0t$ is increasing.
We can now assert that, for $j<(\tau/h)\wedge(n-2)$, ${\cal D}u(B^{(n)}_{jh}, jh)=\tilde{{\cal D}}u(B^{(n)}_{jh}, jh)$,
so that
\begin{eqnarray*}
\E\left(u(\tau,B^{(n)}_\tau)-u(0,0)\right)&\leq &
       \E\left(\sum_{j=1}^{n-2}\left|\tilde{{\cal D}}u(jh,B^{(n)}_{jh})\right|\right)+O(h)\\
       &\leq&
       C\frac{(\ln n)^\alpha}{n},
\end{eqnarray*}
with $\alpha=1$ if $d>r$ and $\alpha=5/4$ if $d\leq r$, as follows from the discussion in the previous section.

We now want a lower bound for 
\[
\E\left(e^{-r\tau}g(\mu _0\tau+B^{(n)}_\tau)-u(\tau,B^{(n)}_\tau)\right).
\]
We have, using the equality $\{\tau\geq T-h\}=\{\tau=T\}$,
\begin{eqnarray*}
u(\tau,B^{(n)}_\tau)-e^{-r\tau}g(\mu_0 \tau+B^{(n)}_\tau)&=&
   \left(u(\tau,B^{(n)}_\tau)-e^{-r\tau}g(\mu_0 \tau+B^{(n)}_\tau)\right)\ind{\{\tau<T-h\}}\\
    &=&\left(u(\tau+h,B^{(n)}_\tau)-e^{-r\tau}g(\mu_0 \tau+B^{(n)}_\tau)\right)\ind{\{\tau<T-h\}}\\
     && +\left(u(\tau,B^{(n)}_\tau)-u(\tau+h,B^{(n)}_\tau)\right)\ind{\{\tau<T-h\}}.
\end{eqnarray*}
On the set $\{\tau<T-h\}$, we have $d(B^{(n)}_\tau, I_{\tau+h})\leq 
                \sqrt{h}||X||_{\infty}+|\mu|h$.
It follows from Proposition~2.6 of \cite{DL2002} that
\[
u(\tau+h,B^{(n)}_\tau)-e^{-r\tau}g(\mu_0 \tau+B^{(n)}_\tau)\leq
C\frac{\left(\sqrt{h}||X||_\infty+|\mu|h\right)^2}{\sqrt{T-\tau-h}}.
  \]
Using  the estimate $\left|\left|\frac{\partial u}{\partial t}(t,.)\right|\right|_\infty<C/\sqrt{T-t}$, we obtain
\[
\E\left(u(\tau,B^{(n)}_\tau)-e^{-r\tau}g(\mu_0 \tau+B^{(n)}_\tau)\right)\leq
     Ch\E\left({1\over \sqrt{T-\tau- h}}\ind{\{\tau\leq T-2h\}}\right).
\]
The estimate $P^{(n)}-P\geq -C\frac{(\ln n)^{\bar{\alpha}}}{n}$ is now an easy consequence
of Lemma~5.7 and Remark 5.8 of \cite{DL2002}, which can be summarized in the following statement.
\begin{lemma}\label{lemma-final}
There exists a positive constant $C$ such that
    \[
    \E\left(\frac{1}{\sqrt{T-\tau- h}}\ind{\{\tau\leq T-2h\}}\right)\leq C\left(\ln h \right)^{\beta},
    \]
    with 
\[\beta=\left\{ \begin{array}{l}
                                                     3/2, \mbox{ if } d\leq r,\\
                                                     \\
                                                     1,  \mbox{ if } d>r.
                                                        \end{array}
                                                         \right.
                                                         \]
\end{lemma}

\end{document}